# Full-Trajectory Learning of Open Quantum Systems: Dynamical Emulation and the Limits of Hamiltonian Identifiability

Ahmad Salmanogli
Department of Electrical and Electronics, Engineering Faculty, Ankara Yildirim Beyazit University, Ankara, Turkey
Corresponding Author email: Ahmadsalmanogli@aybu.edu.tr

**Abstract**
Accurate identification of unknown quantum systems is essential for quantum computing, sensing, and control because the Hamiltonian determines quantum-state evolution. We develop a QNN framework for black-box identification and emulation of open quantum systems using density-matrix trajectories. Rather than matching only final states or selected observables, the QNN is trained through a differentiable open-system propagator to reproduce the full temporal evolution under Lindblad dynamics. A physics-informed synthetic dataset of admissible Hamiltonians and dissipation rates emulates experimental measurements, while the QNN generates a 32-dimensional Hamiltonian coefficient vector used to construct and propagate a Hermitian Hamiltonian. Randomized initial states and chirped modulation of the propagator are utilized as complementary training strategies. Performance is evaluated using weighted trajectory-density loss, quantum-state fidelity, and trace distance. Randomized initialization improves state reconstruction, achieving fidelities of 0.929 for the single-qubit benchmark and 0.787 for the unknown system, with corresponding trace distances of 0.124 and 0.316. Chirped modulation primarily improves optimization, reducing the dominant loss curvature by more than an order of magnitude and decreasing convergence from 16 to 6 epochs for the single-qubit benchmark. The achievable fidelity is bounded by the largest eigenvalue of the target density matrix; calculated fidelities reach 0.61–1.00 of this bound, exceeding 0.90 in 12 of 18 training runs. Sensitivity analysis identifies three exactly null and approximately ten numerically flat parameter directions, explaining non-unique Hamiltonian recovery despite accurate dynamical reconstruction. Finally, the learned Hamiltonian is mapped to a two-qubit–bus-resonator circuit-QED architecture, enabling extraction of transmon capacitances, Josephson inductances, qubit separation, and bus-resonator length. The framework connects black-box quantum-system identification with physical emulation and quantum digital twins while quantifying its identifiability limits.

## Introduction

Quantum technologies have emerged as one of the most transformative research areas in modern science, with broad applications in quantum computing, quantum communication, quantum sensing and quantum simulation [1–4]. The performance of these technologies depends fundamentally on an accurate understanding of the underlying quantum dynamics, which is governed by the system Hamiltonian [5–8]. Because the Hamiltonian completely determines the evolution of a quantum state, its precise identification is essential for device calibration, quantum control, error mitigation and the development of high-fidelity quantum emulators [6–10]. In line with recent developments in quantum computing and artificial intelligence, machine-learning-based Hamiltonian reconstruction and Hamiltonian learning have become rapidly growing research directions. These approaches infer the Hamiltonian of an unknown quantum system directly from experimental observations, enabling efficient system identification, quantum control, digital-twin modelling and quantum simulation without relying on exhaustive quantum state or process tomography [11–15, 41]. Hamiltonian reconstruction has long been recognized as one of the central problems in quantum information science, because the Hamiltonian uniquely determines the coherent evolution of quantum systems through the Schrödinger equation or, in open quantum systems, through the Lindblad master equation [5–8]. Traditionally, Hamiltonian estimation has been performed using quantum state tomography (QST), quantum process tomography (QPT), spectroscopic measurements, Bayesian estimation and maximum-likelihood parameter estimation [7, 16–17]. In quantum state tomography [16], a large number of measurements are performed on identically prepared quantum states to reconstruct the complete density matrix, after which the underlying Hamiltonian parameters can be inferred. Quantum process tomography extends this concept by reconstructing the complete quantum channel that describes the system evolution [24]. Both families of methods have been refined considerably – through direct fidelity estimation from few Pauli measurements [25], rigorous error bars for tomographic estimates [26], compressed-sensing reconstruction of low-rank states [27] and self-guided adaptive protocols [28] – yet their measurement cost still grows unfavorably with system size, which is precisely the bottleneck that learning-based identification seeks to avoid. Alternatively, spectroscopy-based methods estimate Hamiltonian coefficients from measured transition frequencies, Rabi oscillations or Ramsey interference experiments, while Bayesian inference and maximum-likelihood estimation improve parameter estimation by incorporating statistical models and prior information into the reconstruction process [17–18].

Accurate Hamiltonian reconstruction is indispensable across nearly every branch of quantum technology, because the Hamiltonian determines the system dynamics, the energy spectrum and the interaction mechanisms. In quantum computing, precise Hamiltonian estimation enables qubit calibration, gate characterization, quantum error mitigation and the implementation of high-fidelity quantum gates [19–20], and it underpins the reliable execution of quantum algorithms on real hardware [58]. Superconducting quantum processors require continuous identification of qubit frequencies, coupling strengths, anharmonicities and decoherence parameters in order to maintain reliable operation [21]. In quantum sensing and quantum metrology, reconstructed Hamiltonians allow external magnetic fields, electric fields, temperature variations and mechanical perturbations to be inferred with high precision from measured quantum responses. Hamiltonian learning is equally fundamental in quantum control, where optimal control pulses are designed from an accurate physical model of the device [15, 22], and recent work has established that many-body Hamiltonians can in principle be learned with Heisenberg-limited scaling of the total evolution time [41], which sets the theoretical target against which practical schemes should be judged. Furthermore, digital-twin technology for quantum hardware relies on reconstructed Hamiltonians to create virtual replicas capable of predicting device behavior before experimental implementation. Additional applications include quantum simulation, molecular dynamics, condensed-matter physics, quantum communication and the characterization of many-body interacting systems [1–4].

Motivated by recent advances in machine-learning-based Hamiltonian reconstruction and quantum-system identification [5,6,29–35], this work addresses the emulation of an unknown open quantum system whose underlying Hamiltonian is assumed to be completely inaccessible. The objective is that, once the system dynamics have been learned, the behavior of the original quantum system can be reproduced by implementing the reconstructed Hamiltonian in a programmable quantum simulator. Because experimental datasets for genuinely unknown quantum systems are generally unavailable, we first construct a synthetic dataset by randomly generating physically admissible Hamiltonian coefficients and simulating the corresponding open-system dynamics with the Lindblad master equation. This stochastic data-generation framework mimics experimentally measured quantum trajectories while providing sufficient diversity for supervised learning. A quantum neural network (QNN) then produces the Hamiltonian coefficients directly, and the resulting Hermitian operator is propagated by a differentiable open-system solver whose output is compared with the measured trajectory at every sampled time. Rather than estimating quantum states through computationally expensive tomography, the framework therefore inverts the map from Hamiltonian parameters to observed dynamics by differentiating through the dynamics themselves. The estimated Hamiltonian is subsequently translated into an equivalent programmable quantum circuit intended to emulate the dynamics of the original unknown quantum system. Several learning paradigms have been proposed for this purpose, including expectation-value learning based on a chosen observable $\langle a \rangle$, final-state density-matrix ($\rho_f$) learning, inverse-phase-estimation-inspired methods, and complete density-matrix trajectory learning $\rho(t)$ [6,29–35]. The present study adopts complete density-matrix trajectory learning throughout, because it exploits the entire temporal evolution of the quantum state and therefore constrains the model with both transient and steady-state information rather than with a single terminal snapshot. Two complementary training strategies are then examined on top of this common baseline: a chirped modulation of the differentiable propagator and a randomization of the initial quantum state. A central contribution of this work is that these strategies are not only compared empirically but also analyzed quantitatively. We show that they act on different aspects of the problem – the chirp on the conditioning of the optimization landscape, the randomized initialization on the state-level agreement – and we establish the fundamental limits that bound the achievable fidelity, the trace distance and the recoverability of individual Hamiltonian coefficients within the chosen operator basis. This combination of a working emulation pipeline with an explicit identifiability analysis is, to our knowledge, absent from previous Hamiltonian-learning studies and is essential if such models are to be used as quantum digital twins.

## II. Theoretical Framework

### *A. Step-by-step definition and the relevant algorithm*

Fig. 1 illustrates the complete workflow of the proposed machine-learning framework for the identification and emulation of unknown open quantum systems from full density-matrix trajectories. The framework is organized into five consecutive stages that transform randomly generated quantum dynamics into a physically realizable quantum emulator. The first stage is the generation of the unknown quantum system. Since experimentally measured Hamiltonians are generally unavailable for arbitrary quantum devices, the unknown system is modelled by sampling coefficients over a structured 32-operator Hermitian basis together with the corresponding dissipation parameters, namely the resonator decay rate $\kappa$ and the two qubit relaxation rates $\gamma_1$ and $\gamma_2$. These parameters define a completely unknown open quantum system governed by the Lindblad master equation. Random generation allows the framework to cover a broad family of experimentally realizable superconducting quantum circuits rather than a single predefined model. The second stage performs dataset generation. The unknown Hamiltonian is propagated with the Lindblad equation using QuTiP, producing the complete density-matrix evolution over the entire time interval. Instead of recording only the final quantum state, every density matrix along the trajectory is stored, so that each sample contains the complete temporal dynamics of the system, including transient coherent oscillations and dissipative relaxation. Gaussian measurement noise is added to each stored density matrix to emulate realistic experimental conditions. The third stage is the QNN module. A deep feed-forward network receives a compact control vector – optionally concatenated with the description of the randomized initial state – and outputs thirty-two coefficients over the predefined operator basis. These coefficients are assembled into a Hamiltonian that is explicitly symmetrized so that the Hermiticity requirement of quantum mechanics is satisfied. The predicted Hamiltonian then enters a differentiable open-system evolution module that reproduces the full quantum trajectory under the same dissipation parameters as the target. The fourth stage performs trajectory-based optimization. The predicted density matrices are compared with the target density matrices at every sampled time instant rather than only at the final evolution time, and a weighted trajectory loss accumulates the reconstruction error throughout the evolution, so that short-time and long-time dynamics are learned simultaneously. Gradients are backpropagated through the differentiable evolution and all network parameters are updated with the Adam optimizer until convergence. Within this stage, two complementary strategies are available: a chirped modulation α(t) applied to the predicted effective Hamiltonian, and a randomization of the initial quantum state used for both the target and the prediction. Exposing the optimizer to a diverse set of initial conditions broadens the range of quantum evolutions that the model must reproduce and reduces the bias towards a single state preparation [36–38]. The final stage is the construction of the quantum emulator. Once optimization converges, the estimated coefficients constitute a complete mathematical representation of the identified open quantum system, which is projected onto a two-qubit–bus-resonator circuit-QED architecture in the dispersive regime. Standard transmon and coplanar-waveguide relations then convert the identified frequencies, couplings and decay rates into circuit-level quantities – total qubit capacitances, Josephson inductances, an equivalent inter-qubit coupling capacitance, the bus-resonator length and the qubit–qubit separation – which together define a physically realizable emulator, or digital twin, of the unknown system.

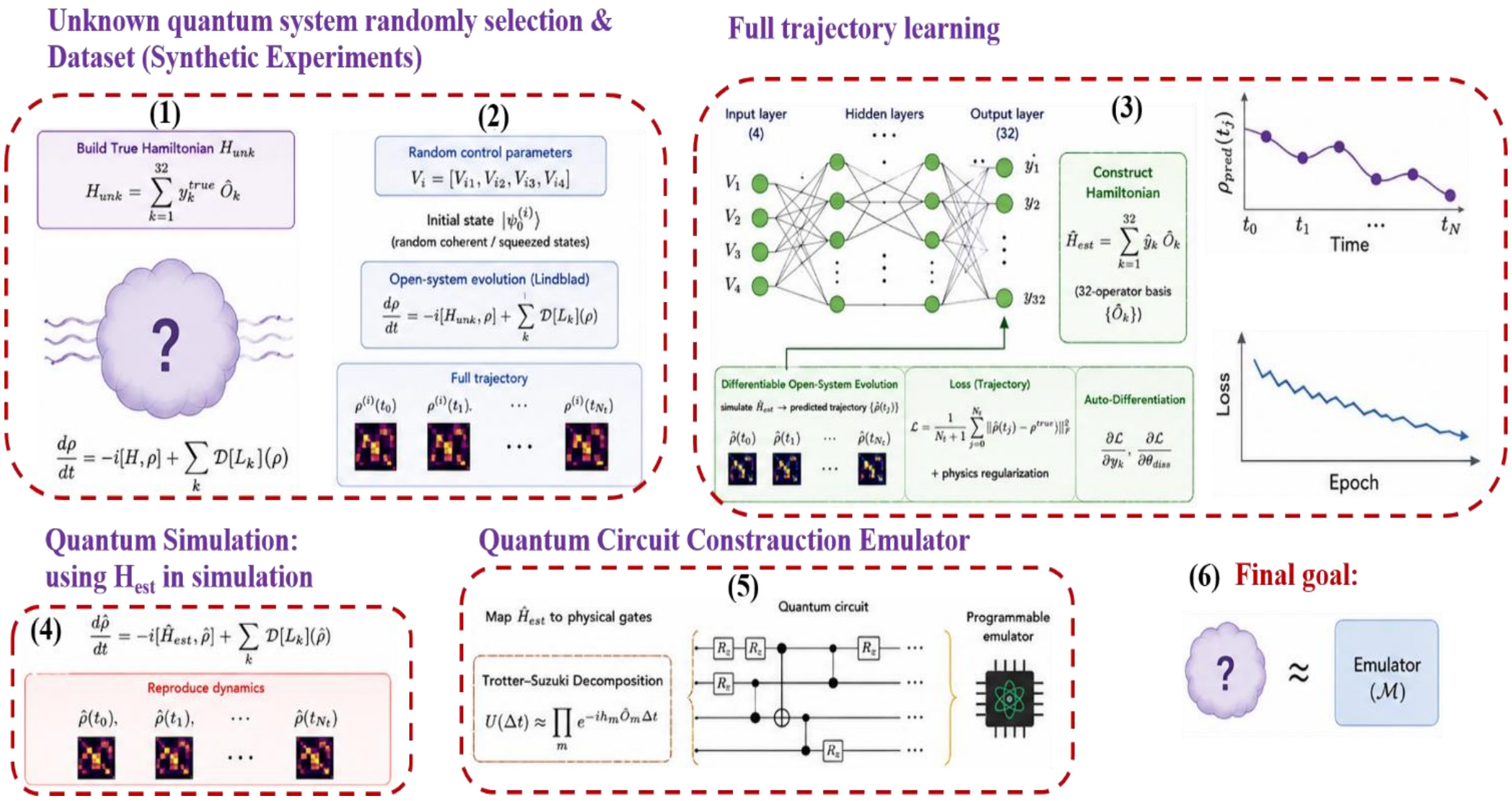

Fig. 1. Overall workflow of the proposed full-trajectory QNN framework for the identification of an unknown quantum system and the construction of a quantum emulator. (1) An unknown open quantum system is generated by physics-informed random sampling of coefficients over a structured 32-operator basis together with the dissipation rates κ, $\gamma_1$ and $\gamma_2$. (2) The corresponding Lindblad dynamics are propagated with QuTiP, and the complete noisy density-matrix trajectory is stored together with the control vector and, where used, the randomized initial state. (3) A five-layer QNN maps the control vector onto the thirty-two Hamiltonian coefficients, from which a Hermitian Hamiltonian is assembled. (4) The predicted Hamiltonian is propagated by a differentiable open-system solver, optionally under a chirped modulation α(t), and the weighted trajectory loss is minimized with the Adam optimizer through backpropagation. (5) The converged coefficients are projected onto a two-qubit–bus-resonator cQED architecture, which provides a physically realizable emulator, or digital twin, of the unknown system.

Algorithm 1. Physics-Informed Quantum Neural Network for Unknown Hamiltonian Learning

**Input:** Hamiltonian basis B={B1,...,B32},initial density matrix ρ0, time sequence T,dissipation ranges {κ,γ1,γ2},Nsamples, Nepochs.
**Output:** Optimized Hamiltonian coefficients y* and reconstructed Hamiltonian Hlearned.

function QuantumHamiltonianLearning()

**#Dataset Generation**

**for** i = 1 → $N_{samples}$ **do**

Randomly generate hidden coefficients $\theta_{hidden}$

Construct $H_{unknown} = \Sigma\ \theta_{hidden}(k)B_k$

Sample $\kappa$, $\gamma_1$, $\gamma_2$

Solve Lindblad master equation $d\rho/dt= -i[H_{unknown},\rho]+ \Sigma\ D[L_k](\rho)$

Obtain ρ(t1),ρ(t2),...,ρ(tN)

Add measurement noise

Generate random control vector

$V=[V_1,V_2,V_3,V_4]$

Store (V, density trajectory, $\theta_{hidden}$, dissipation)

**end for**

**#Neural Network Training** #Initialize QNN parameters

**for** epoch = 1 → $N_{epochs}$ **do**

**for** each mini-batch **do**

Predict y = QNN(V)

Construct $H_{pred} = \Sigma\ y_k B_k$

Enforce Hermiticity $H_{pred} \leftarrow (H_{pred} + H_{pred}^{\dagger})/2$

Propagate the predicted state

ψ(t) ← differentiable no-jump evolution under α(t)Heff

for every sampled time step tj do

Form ρpred(tj) = ψ(tj)ψ(tj)†

Accumulate the weighted density loss

$wj\ ||\rho_{pred}(t_j) - \rho_{meas}(t_j)||^2 / (N\ D^2)$

**end for**

Normalize the weighted trajectory loss by N

Backpropagate gradients

Update QNN parameters using Adam

**end for**

**end for**

**#Hamiltonian Reconstruction**

Obtain optimized coefficients y*

Construct $H_{learned} = \Sigma\ y^*kB_k$

Return y*, $H_{learned}$

**end function**

***B. Theoretical backgrounds***

The learning procedure begins with an unknown open quantum system whose dynamics are governed by the Lindblad master equation [43, 54-57, 65, 66] expressed as:

$$\frac{d\rho(t)}{dt} = -i[H_{\text{unknown}}, \rho(t)] + \sum_k \mathcal{D}[L_k]\rho(t), \quad (1)$$

where *ρ(t)* denotes the density matrix describing the quantum state, $H_{unkown}$ is the unknown Hamiltonian, and $L_k$ are Lindblad collapse operators representing dissipation channels such as resonator decay and qubit relaxation presented generally as [43, 54-57]:

$$\mathcal{D}[L]\rho = L\rho L^\dagger - \frac{1}{2}(L^\dagger L\rho + \rho L^\dagger L) \quad (2)$$

The unknown Hamiltonian is represented by [42]:

$$H_{\text{unknown}} = \sum_{k=1}^{32} \theta_k\, O_k, \quad (3)$$

where $O_k$ are predefined Hermitian basis operators and $\theta_k$ are the coefficients to be learned. The basis used throughout this work contains thirty-two operators acting on the sixteen-dimensional Hilbert space of two qubits and two two-level bosonic modes, and is organized into physically meaningful groups: single-qubit operators ($\sigma_x$, $\sigma_z$ on each qubit), bosonic number operators and their squares, quadrature operators, dispersive qubit–mode couplings, transverse light–matter couplings, two-qubit XX, YY and ZZ interactions, cross-Pauli terms and the resonator-mediated exchange interaction. To emulate the absence of prior physical knowledge, the coefficients of the unknown system are sampled as $\theta_k \sim R(-1, 1) \times 2\pi$ GHz,, forming the coefficient vector $\theta = [\theta_1, \theta_2, \ldots, \theta_{32}]$. Here $R(\cdot)$ denotes a physics-informed rather than a uniform sampling rule: although the coefficients themselves are drawn at random, the operators are pre-weighted so that the classes that dominate practical superconducting hardware – qubit energies, resonator energies and dispersive couplings – enter with amplitudes of order unity, whereas higher-order interactions, quadrature excitations, nonlinear bosonic terms and cross-Pauli couplings are scaled by factors between 0.01 and 0.15. The generated Hamiltonians therefore retain the stochastic diversity required for a broad training set while remaining consistent with the interaction hierarchy of real circuit-QED devices, and the network is not trained on arbitrary Hermitian matrices that would be unlikely to occur experimentally. Two properties of this parameterization should be stated explicitly, because they govern the interpretation of every result reported below. First, the thirty-two operators are not linearly independent: the Hilbert–Schmidt Gram matrix of the basis has rank twenty-nine. The exchange operator $\sigma_+^{(1)}\sigma_-^{(2)}$ + h.c. is exactly the mean of the XX and YY operators, and, because both bosonic modes are truncated to their lowest two levels, the Kerr-type operators like $na^2$ and $nb^2$ coincide identically with the number operators $na$ and $nb$. Three coefficient combinations therefore leave the Hamiltonian – and hence every observable – strictly unchanged. Second, the effective number of coefficients that the measured dynamics can constrain is smaller still, as quantified in Sec. III. These structural degeneracies are a property of the operator basis, not of the neural network, and they are the reason why the framework is presented here as a route to dynamical emulation rather than to unique operator-level identification. Each realization corresponds to a different quantum system with its own interactions, frequencies and coupling strengths. The generated Hamiltonian is propagated to produce the complete density-matrix trajectory $D = \{V, \rho(t_1), \rho(t_2), \ldots, \rho(t_N)\}$, where V is the control vector [42] and $\rho(t_i)$ is the density matrix at the $i^{th}$ of N = 200-time instants spanning 50 ns. To mimic experimental measurement, two independent Gaussian terms with standard deviations $1\times10^{-4}$ and $2\times10^{-4}$ are added to every stored density matrix, $\rho_{noisy} = \rho + N(0, \sigma^2)$ with an effective $\sigma = 2.2\times10^{-4}$. The dissipation parameters are drawn independently for each realization, with $\kappa/2\pi \in [8, 33]$ MHz and $\gamma_1/2\pi$, $\gamma_2/2\pi \in [0.1, 1]$ MHz, and are supplied to both the target and the predicted evolution so that the learning problem concerns the Hamiltonian alone. The QNN learns a nonlinear mapping from its input to the Hamiltonian coefficients, $y = f_{QNN}(V)$, where $f_{QNN}(\cdot)$ denotes the neural network.

The objective of the QNN in this study (Fig. 1) is not to predict the quantum state directly; the network instead generates the Hamiltonian coefficients from which the state is obtained by explicit physical propagation. This is what makes the framework physics-informed: the neural network supplies a smooth, differentiable parameterization of an operator, while the quantum mechanics is carried by the Lindblad propagator rather than learned from data. Two network variants are used, matching the two initial-state strategies. In the fixed-initial-state configuration the network receives the four-dimensional control vector $V = [V_1, V_2, V_3, V_4]$ and consists of five fully connected layers arranged as $4 \rightarrow 16 \rightarrow 32 \rightarrow 128 \rightarrow 64 \rightarrow 32$. The first two hidden layers use the hyperbolic-tangent (*tanh*) activation, which captures nonlinear relations between the control variables and the dynamics while maintaining smooth gradients during backpropagation; the third layer expands the representation to 128 neurons with a *Sigmoid* activation, providing a high-capacity feature space for the correlations that arise in open-system evolution; the fourth layer compresses the representation to 64 neurons with a *tanh* activation; and the linear output layer produces the thirty-two coefficients. In the randomized-initial-state configuration, the network additionally receives the real and imaginary parts of the sampled initial state vector, so that the input dimension becomes $4 + 2D = 68$ with $D = 16$, and the layer widths are enlarged to $68 \rightarrow 64 \rightarrow 128 \rightarrow 256 \rightarrow 128 \rightarrow 32$ with the same activation pattern. In both cases the raw output is passed through a hyperbolic tangent and rescaled, $y = \tanh(f_{QNN}(\cdot)) \times 2\pi$ GHz, which confines every coefficient to a physically sensible band and keeps the propagator numerically well conditioned throughout training. The predicted coefficients are then combined with the predefined 32-operator basis to reconstruct the estimated Hamiltonian, which is explicitly symmetrized to satisfy the Hermiticity condition required for a physically realizable quantum system. This reconstructed Hamiltonian is subsequently used inside the differentiable open-system evolution module to generate the predicted quantum trajectory, and the discrepancy between the predicted and target density-matrix trajectories is minimized by gradient-based optimization [34–35, 42]. It should be noted that, in the present implementation, the control vector V is drawn independently of the system that generated the trajectory with which it is paired. The map $V \mapsto y$ is therefore an over-parameterized, smoothly regularized generator of a coefficient vector – comparable in spirit to an implicit neural parameterization – rather than a predictive regressor that infers a Hamiltonian from measured control settings. The consequences of this choice are quantified in Sec. III and revisited when the limitations of the framework are discussed. The estimated coefficients are assembled into the predicted Hamiltonian exactly as in Eq. 3, and the resulting operator is propagated by a differentiable quantum evolution module implemented entirely in PyTorch [45–47] as:

$$\hat{H} \leftarrow \frac{1}{2}(\hat{H} + \hat{H}^{\dagger}), \tag{4}$$

where $\hat{H}^{\dagger}$ denotes the conjugate transpose, which guarantees a physically admissible generator. The reconstructed Hamiltonian is then propagated by the differentiable evolution [54]:

$$\psi(t + \Delta t) = e^{-iH_{\mathrm{tot}}(t)\Delta t}\psi(t), \tag{5}$$

where $H_{tot}(t) = \alpha(t)H_{eff}$, $\Delta_t$ is the integration time step, $H_{eff} = H_{pred} - (i/2)(\kappa\, a^{+}a + \gamma_1\, \sigma_+^{(1)}\sigma_-^{(1)} + \gamma 2\, \sigma_+^{(2)}\sigma_-^{(2)})$ is the effective non-Hermitian Hamiltonian that carries the dissipation, and *α(t)* is an optional chirped modulation. Two features of Eq. 5 must be stated precisely, because both are essential to the interpretation of the results. First, Eq. 5 is a no-jump, norm-renormalized quantum-trajectory propagator: the anti-Hermitian part of $H_{eff}$ reproduces the deterministic damping of the Lindblad generator, but the recycling term that repopulates the lower states after a quantum jump is not included, and the state is renormalized after every step. The predicted state is consequently always pure, $\rho_{pred}(t) = |\psi(t)\rangle\langle\psi(t)|$, whereas the target trajectory obtained from the full Lindblad solver becomes mixed as the evolution proceeds. This makes the propagator inexpensive and exactly differentiable, which is what allows back-propagation through two-hundred-time steps, but it also imposes a hard limit on the achievable state-level agreement: for a pure prediction the *Uhlmann fidelity* [49, 50, 61, 62] reduces to $F = \langle\psi|\rho_{unk}|\psi\rangle$ and is therefore bounded above by the largest eigenvalue $\lambda_{max(\rho unk)}$, while the trace distance obeys $D \geq 1 - \lambda_{max}$. These bounds are evaluated for every run in Sec. III and are found to account for most of the residual discrepancy. Second, the modulation $\alpha(t)$ multiplies the predicted effective Hamiltonian only; the target trajectories are generated once, without modulation, by the Lindblad solver. The chirp is thus a property of the learner, not an excitation physically applied to the unknown device, and it cannot by itself increase the information that the data carry about the unknown Hamiltonian. What it can do – and, as shown in Sec. III, what it demonstrably does – is reshape the optimization landscape. Two modulation profiles are examined. The sinusoidal chirp $\alpha(t) = A \sin[2\pi(f_0\tau + (f_1 - f_0)\tau^2/2)]$ with $\tau = t/T$ sweeps the effective time scale of the propagator continuously across the evolution window, with $A \in [1.0, 2.25]$, $f_0$ and $f_1$ are selected rationally once per run. The Gaussian profile $\alpha(t) = \exp[-(\tau - 0.5)^2/2\sigma c^2]$ with $\sigma_c = 0.15$ instead concentrates the effective propagation near the center of the window. Because a sign-alternating $\alpha(t)$ reduces the phase accumulated per step and averages the curvature of the loss over a range of effective time scales, the chirped propagator behaves as an implicit preconditioner: it lowers the dominant curvature of the trajectory loss by more than an order of magnitude and mildly compresses its eigenvalue spectrum, which accelerates and stabilizes convergence under a fixed learning rate. This mechanism, which is quantified in Fig. 5, is consistent with the empirical finding reported below that chirped modulation improves the optimization but not the state-level fidelity (based on the finding of this work), and it differs from the information-theoretic interpretation of chirped probing that applies when the chirp is applied to the physical device itself [47–48]. In addition to the chirped propagator, a second training strategy is used to improve convergence, optimization stability and generalization during Hamiltonian reconstruction. Instead of employing a single fixed initial quantum state throughout training, a different initial state is generated for each realization of the unknown quantum system [36–38]. This enriches the diversity of the training set by exposing the QNN to evolutions originating from different regions of the Hilbert space, so that the model is constrained by intrinsic dynamical features of the underlying Hamiltonian rather than by trajectory-specific features tied to one particular state preparation. Similar data-diversification strategies have proven effective in improving the robustness and generalization of machine-learning models for quantum-state reconstruction, Hamiltonian learning and quantum control [36–40]. In the present implementation the two qubits are prepared in the fixed product state $|q_1, q_2\rangle = |1, 0\rangle$, while the two bosonic modes are prepared in coherent states whose amplitudes are re-sampled for every realization, so that the initial state reads $|\Psi_0(i)\rangle = |q_1, q_2\rangle \otimes |\beta_i\rangle \otimes |\alpha_i\rangle$,, where $\alpha_i$ and $\beta_i$ are drawn independently from the uniform distribution on [0.2, 1.0]; for reference, the fixed-initial-state runs use $\alpha = \beta = 0.7$. Because the randomized amplitudes must be known to the model, the description of the initial state is concatenated with the control vector at the network input, which is why the randomized-initialization runs use the larger of the two architectures described above. The dataset can then be written as:

$$\mathcal{D} = \left\{\left(V_i, \psi_0^{(i)}, \rho_i(t), \theta_i\right)\right\}_{i=1}^{N}, \tag{6}$$

where $V_i$ is the control input and $\Psi_0^{(i)}$ is the randomized initial state. The essential point is that the framework adopts a full-trajectory learning strategy for the identification of an unknown, black-box open quantum system [42]: the objective is not to reproduce the final quantum state but to reconstruct the complete dynamical evolution of the density matrix under realistic decoherence. Random Hamiltonian coefficients together with randomly sampled dissipation rates generate a population of synthetic open quantum systems, each of which is simulated with the QuTiP Lindblad master-equation solver [43], producing the sequence of density matrices that constitutes the target trajectory. Gaussian noise [43–44] is added to every stored density matrix so that the network is trained under experimentally realistic measurement imperfections. Instead of comparing only the final quantum state, the method therefore minimizes the discrepancy accumulated over the complete trajectory [42–43]:

$$L = \frac{1}{ND^2}\sum_{j=1}^{N} w_j \parallel \rho_j^{\text{pred}} - \rho_j^{\text{unknown}} \parallel_F^2, \tag{7}$$

where N = 200 is the number of sampled time steps, D = 16 is the Hilbert-space dimension, $||\cdot||_F$ denotes the Frobenius norm [64] and $w_j = j/N$ is a temporal weighting coefficient that increases linearly along the trajectory, so that the accumulated long-time dynamics – which carry the dissipative signatures – are weighted more strongly than the first few steps. The normalization by $N$ and by the $D^2 = 256$ matrix elements makes $L$ the mean squared modulus of the element-wise density-matrix error, which is the form used throughout Sec. III; the equivalent relative error is $||\Delta_\rho||_F / ||\rho_{unk}||_F = (D^2L)^{0.5}/ ||\rho_{unk}||_F$, evaluated at a single time step. In the present implementation only this density term is active. Auxiliary quadrature, variance, photon-statistics and trace-preservation terms are available in the loss function but are assigned zero weight, so that the reported optimization is driven purely by the trajectory-density discrepancy; no observable-specific penalty is applied. The strategy exploits the entire dynamical evolution and provides substantially more information than terminal-state learning. The neural-network parameters are updated with the Adam optimizer [42]:

$$\theta \leftarrow \theta - \eta\nabla_\theta L, \ (8)$$

where $\eta = 10^{-3}$ is the learning rate, $L$ is the trajectory loss and $\nabla_\theta L$ denotes the gradient obtained by automatic differentiation through the two hundred propagation steps. Each model is trained for 100 epochs on 300 sampled systems with a mini-batch size of 32. Iterative optimization minimizes the discrepancy between the predicted and the measured trajectories, so that after convergence $\rho^{pred} \sim \rho^{unknown}$; the reconstructed density matrix then captures the dynamics of the unknown quantum system and can be used for subsequent analysis or implementation. It is worth noting at this stage that the injected measurement noise contributes a loss floor of only $\sigma^2 \approx 5\times10^{-8}$, which is more than three orders of magnitude below the residual losses of $1.5\times10^{-4}$ to $2.2\times10^{-3}$ reported in Sec. III. The residual therefore reflects model and optimization error rather than measurement noise, and the noise level used here should be regarded as a realism device rather than as a test of noise robustness.

The ultimate objective of this study extends beyond the accurate identification of the unknown system Hamiltonian to the reconstruction of a corresponding physical quantum circuit capable of reproducing the observed system dynamics. Once the QNN converges and the predicted density matrix closely matches that of the unknown quantum system, the learned parameters can be translated into physically realizable circuit-design parameters through established circuit quantum electrodynamics (cQED) relationships [54]. In this work, a two-qubit system coupled through a bus resonator is considered as the initial circuit architecture for this reconstruction. The underlying assumption is that the learned density matrix and its associated Hamiltonian can be represented, at least partially, by an equivalent quantum chip (in this work, for the sake of simplicity an optional two-qubit–bus-resonator architecture is supposed as a quantum chip). Consequently, the reconstructed circuit is not necessarily expected to reproduce every microscopic property of the unknown quantum system, but rather to emulate its relevant dynamical behavior within the considered operating regime. If the reconstructed circuit can reproduce the key observables and dynamical characteristics of the unknown system with sufficient fidelity, it can serve as a physics-informed quantum digital twin. Such a digital twin would provide a physically interpretable circuit-level representation of the unknown system and could subsequently be used for simulation, prediction, parameter exploration, and circuit-level optimization without requiring direct access to the original quantum device. In the following, we attempt to concentrate on the simulation results and discuss the different aspects of the approach utilized in this study.

**Results and Discussions**

In this work we develop a comprehensive QNN-based framework for the emulation of unknown quantum systems through full-trajectory learning. Rather than limiting the investigation to a single physical model, we first consider two representative systems with explicitly known Hamiltonians as benchmark cases for validating the learning framework. The first is a single-qubit Hamiltonian, $H_1 = 0.5\times\omega_1\sigma_z^1$ with $\omega_1/2\pi$ drawn from [4.9, 5.1] GHz, while the second describes two qubits coupled through a bus resonator, $H_2 = 0.5\times\omega_1\sigma_z^{(1)} + 0.5\times\omega_2\sigma_z^{(2)} + (\omega_r + \chi_1\,\sigma_z^{(1)} + \chi_2\sigma_z^{(2)})a^{+}a + J_{12}\times(\sigma_-^{(1)}\sigma_+^{(2)} + \sigma_-^{(2)}\sigma_+^{(1)})$, with $\omega_1/2\pi \in [4.9, 5.1]$ GHz, $\omega_2/2\pi \in [5.1, 5.3]$ GHz, $\omega_r/2\pi \in [6.0, 7.2]$ GHz, $\chi/2\pi \in [1, 20]$ MHz and $J_{12}/2\pi \in [5, 100]$ MHz. The third case is the genuinely unknown system, for which all thirty-two coefficients of the operator basis are sampled independently. These three cases represent progressively increasing levels of physical complexity and provide a controlled means of assessing the framework before it is applied to an inaccessible Hamiltonian. The learning performance is evaluated primarily through the trajectory-density loss $L_\rho$ of Eq. 7, which quantifies the discrepancy between the predicted and target density-matrix trajectories over the complete evolution. For all investigated systems, full-trajectory learning is the common training strategy, and the complementary techniques introduced in Sec. II are applied on top of it: chirped modulation of the differentiable propagator, in a sinusoidal and a Gaussian variant [47–48], and randomized initial conditions [36–38]. The resulting strategies (around six) are compared according to three criteria: the rate of convergence towards the minimum, the converged value of the trajectory-density loss, and the magnitude of the residual fluctuations around it. Unless stated otherwise, every run uses 300 sampled systems, a mini-batch size of 32, 100 training epochs, a learning rate of $10^{-3}$, and a trajectory of 200 density matrices spanning 50 ns; the dissipation rates are resampled for every system as specified in Sec. II. Increasing the number of training samples would be expected to improve accuracy further, but the present dataset size was selected according to the available computational resources. Two conventions should be kept in mind when reading the figures and tables that follow. The loss curves of Figs. 2–4 report the population-average loss over the whole training set at the end of each epoch, whereas the fidelity and trace distance of Table III are evaluated, as in the original implementation, for the single best-matching realization encountered during training. The two metrics therefore address distinct questions: the first assesses how accurately the model fits the overall ensemble, whereas the second evaluates how well the best identified Hamiltonian reproduces a specific unknown system. Accordingly, they are treated as separate analyses and presented independently in Figs. 5 and 8.

Fig. 2 compares the influence of several learning strategies on the convergence behavior of the proposed QNN framework, including sinusoidal chirp excitation, Gaussian chirp excitation, randomization of the initial conditions, and combinations of these techniques. The results are compared with the baseline approach without additional excitation or initialization strategies. Although the combined approaches exhibit a relatively rapid initial reduction in the trajectory-density loss, their final steady-state loss remains comparatively high, indicating that fast convergence does not necessarily guarantee accurate reconstruction. It should be noted that we applied the combination of chirped excitation and randomized initial conditions to three different Hamiltonians and consistently observed the same behavior: a relatively rapid initial reduction in the trajectory-density loss, followed by a comparatively high steady-state loss. However, this observation is based on a limited set of Hamiltonians and should not be interpreted as a general or definitive rule. Further investigation across a broader range of Hamiltonian structures and system parameters is required to establish whether this behavior is systematic. In contrast, the baseline approach, represented by the black curve, exhibits considerably slower convergence, but a fair final steady-state loss. Among the investigated strategies, the individual sinusoidal chirp excitation and randomized initial conditions provide the most favorable overall behavior, achieving both a relatively rapid convergence and a lower final trajectory-density loss. This result suggests that appropriately designed excitation can improve the richness of the observed quantum dynamics and help the QNN explore a broader region of the Hamiltonian parameter space. Accordingly, the sinusoidal-chirp and randomized-initial-condition strategies are adopted for the subsequent Hamiltonian-learning presented in Fig. 3.

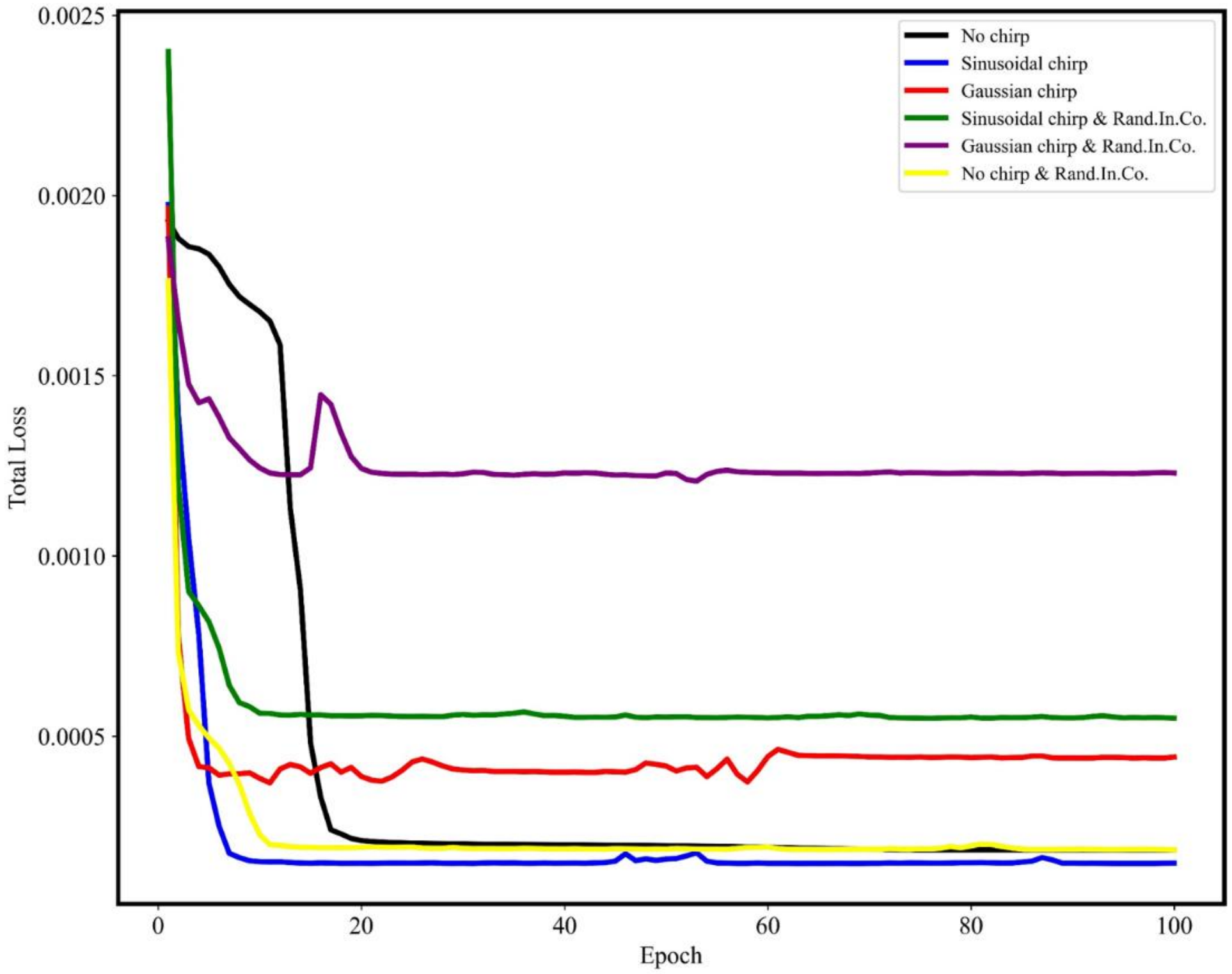

Fig. 2 Trajectory-density loss achieved for different approaches for the case of single qubit Hamiltonian $H_1 = 0.5\times\omega_1\sigma_{z1}$; “Rand.In.Co.” is acronym for randomized initial conditions.

Fig. 3 further demonstrates the influence of Hamiltonian complexity on the reconstruction performance. As the number of effective degrees of freedom increases, the trajectory-density loss becomes larger, indicating that the learning problem becomes progressively more difficult. This is particularly evident for the unknown Hamiltonian, which is represented by thirty-two independent coefficients, whereas the benchmark Hamiltonians $H_1$ and $H_2$ contain substantially fewer effective degrees of freedom and are therefore learned more accurately from the available training data. As shown quantitatively in Fig. 5, the degradation observed for the unknown system does not indicate a limitation of the QNN architecture itself; it reflects two distinct effects acting together. The first is the increased dimensionality and the correspondingly larger parameter-identification burden of a highly flexible Hamiltonian representation. The second, and in the present setting the dominant one, is that the training ensemble for the unknown case consists of mutually unrelated random Hamiltonians, so that no single coefficient vector can serve the population. In the present simulations 300 sampled systems and 100 training epochs were employed; increasing the number and, more importantly, the structural coherence of the training trajectories would be expected to improve parameter identifiability and to reduce the residual reconstruction error. Nevertheless, the results demonstrate that the framework reconstructs the dynamics of systems of increasing complexity and provides a practical basis for extending the approach towards genuinely unknown quantum systems. Consistent with the observations in Fig. 2, the different learning strategies were subsequently applied to the genuinely unknown Hamiltonian, and the results presented in Fig. 4 confirm the trends identified in the preliminary comparison. In particular, the combination of full-trajectory learning with sinusoidal-chirp modulation provides the most favorable overall optimization behavior for the identification of the black-box quantum system, reaching both the lowest plateau loss and the largest fraction of loss removed among the strategies examined for that case.

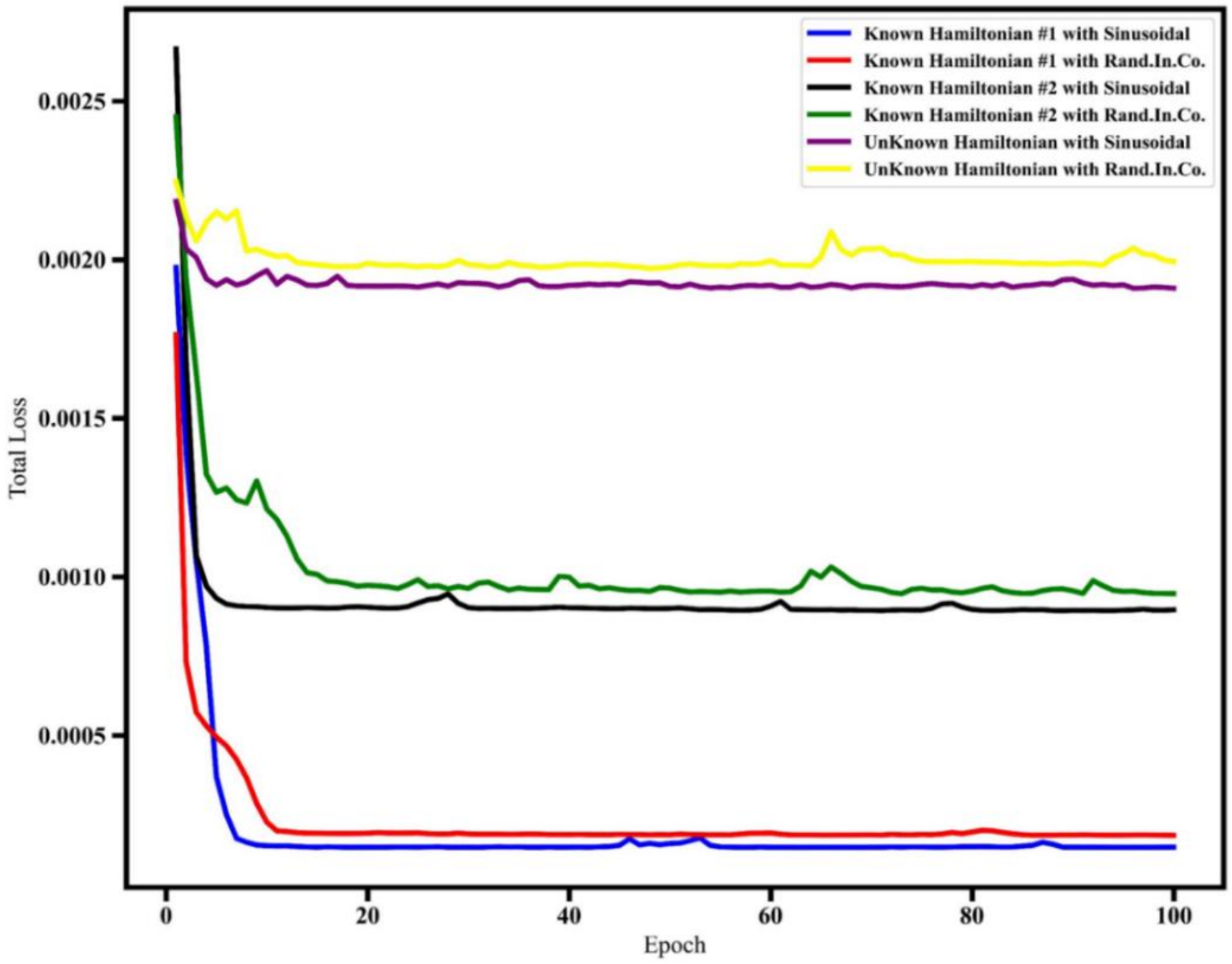

Fig. 3 Trajectory-density loss achieved for different approaches applied on different Hamiltonians

As established in Sec. II and quantified in Fig. 5, the benefit of the chirp is one of conditioning rather than of information: sweeping the effective time scale of the differentiable propagator flattens the dominant curvature of the trajectory loss and thereby allows the optimizer to descend further within the same number of epochs, while the full-trajectory objective supplies the dynamical information by constraining the model at every sampled time instead of only at the final state. A further point concerns the magnitude of the fluctuations around the converged solution in Fig. 4. These fluctuations can be attributed to the random selection of the Hamiltonian coefficients across the training ensemble, which roughens the optimization landscape and makes the identification of any single target Hamiltonian more difficult; their magnitude, quantified in Fig. 5c, nevertheless remains below $1.5\times10^{-5}$ for the unknown system, so the plateaus are genuinely converged. Fig. 2–Fig. 4 compare the training strategies qualitatively. To place that comparison on a quantitative footing, and to identify the mechanism through which chirped modulation acts, Fig. 5 reports three convergence descriptors together with a direct analysis of the optimization landscape.

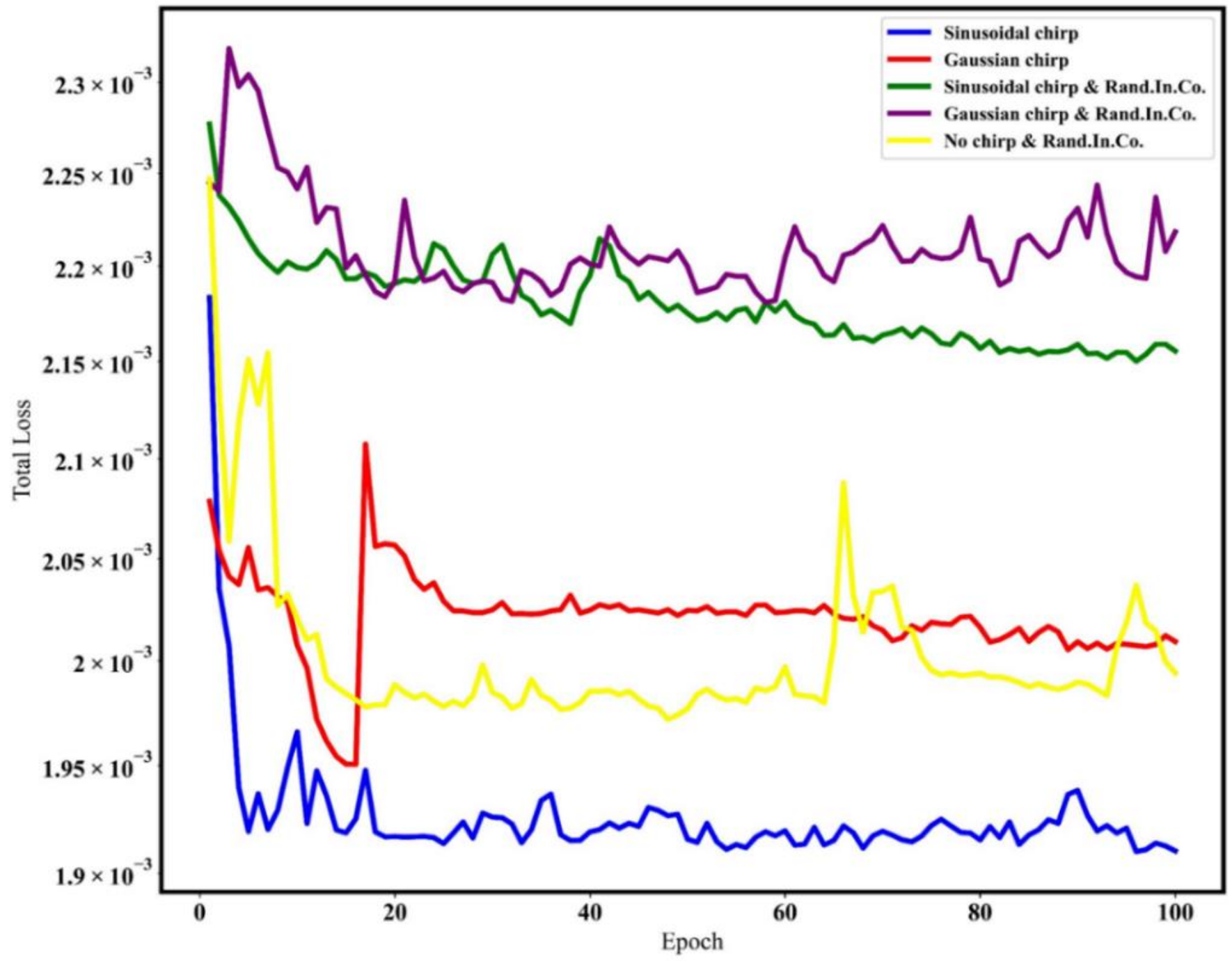

Fig. 4 Trajectory-density loss achieved for different approaches for the case of Unknown Hamiltonian ($H_{unk}$).

Fig. 5a–Fig. 5c give, for every system and strategy, the number of epochs required to realize 90% of the total loss reduction, the fraction of the initial loss that training actually removes, and the converged plateau together with its standard deviation over the final twenty epochs. For the single-qubit benchmark the sinusoidal chirp shortens convergence from sixteen epochs to six while simultaneously lowering the plateau from $1.85\times10^{-4}$ to $1.48\times10^{-4}$, whereas the Gaussian chirp converges fastest, in three epochs, but settles at a threefold higher loss of $4.4\times10^{-4}$. The same ordering holds for $H_2$, where the sinusoidal chirp reaches its plateau in three epochs at $9.0\times10^{-4}$ while the Gaussian variant needs fifteen epochs and stops at $2.0\times10^{-3}$. These numbers substantiate, and make reproducible, the qualitative statement that an appropriately designed modulation improves both the speed and the quality of convergence, and they also quantify the fluctuation criterion: the residual standard deviation over the last twenty epochs never exceeds $7.5\times10^{-5}$ and is typically below $10^{-5}$, so the plateaus in Figs. 2–4 are genuinely converged rather than slowly drifting. Fig. 5b exposes a feature that the logarithmic loss curves conceal. The fraction of the loss that training removes reaches 90.4% for $H_1$ without modulation and 92.5% with the sinusoidal chirp, but no strategy exceeds 66.4% for $H_2$, and for the 32-coefficient unknown system the best value is 12.5% while three of the six strategies remove less than 4%. This behavior follows directly from the construction of the training set rather than from any limitation of the network. Each sample pairs a different randomly drawn system with an independently drawn control vector, so the input carries no information about which Hamiltonian generated the trajectory with which it is paired; the best solution the mapping can reach is therefore the single coefficient vector that reproduces the ensemble on average. When the ensemble is narrow – $H_1$ varies only through a qubit frequency within a 200 MHz window – that ensemble optimum lies close to every member and the loss falls by almost an order of magnitude. When the ensemble consists of mutually

unrelated random Hamiltonians, no single vector serves the population and the achievable population-level reduction is correspondingly small. This is the correct reading of the trend already visible in Fig. 3, where the trajectory-density loss grows with Hamiltonian complexity: the growth reflects the diversity of the sampled ensemble as much as the dimensionality of the parameterization, and it is the reason why the state-level metrics of Table III are reported for the best-matching individual realization. Fig. 5d–5f analyze the landscape itself, in which the Gauss–Newton (Fisher) metric [73–77] of the weighted trajectory loss is evaluated, $G_{kl} = (1/N)\, \Sigma_j\, w_j\, \mathrm{Re}\{\mathrm{Tr}[(\partial\rho(t_j)/\partial\theta_k)^* \,\partial\rho(t_j)/\partial\theta_l]\}$, at the converged coefficients of the unknown system, using the same differentiable propagator that is used for training. The metric is defined here with the Hilbert–Schmidt trace, so that it is related to the curvature of the loss of Eq. 7 by $\partial^2\mathscr{L}/\partial\theta_k\partial_{\theta l} \approx 2G_{kl}/D^2$; all statements below concern ratios of eigenvalues and are unaffected by this convention, which is developed in Appendix G. The relevant eigenvalue spectrum is shown in Fig. 5d, which is extremely anisotropic. Three eigenvalues vanish identically, in agreement with the exact rank deficiency of the operator basis established in Sec. II, and the twenty-nine remaining eigenvalues span more than seven orders of magnitude, with only about twenty exceeding $10^{-6}$ of the largest. Fig. 5e translates this into directly interpretable terms: displacing the converged coefficients by a full $2\pi\times1$ GHz – the entire output range of the network – along the stiffest eigendirection more than doubles the trajectory loss, whereas the same displacement along the sloppiest identifiable direction changes it by less than $2\times10^{-4}$ in relative terms, and along an exactly null direction leaves it unchanged to machine precision. The learned model is therefore tightly constrained in a handful of collective directions and essentially unconstrained in the remainder, which is the classic signature of a sloppy inverse problem. Finally, Fig. 5f isolates the role of the chirp. Relative to constant propagation, the sinusoidal chirp lowers the dominant curvature $\lambda_1$ from $1.26\times10^3$ to $8.8\times10^1$ and compresses the eigenvalue spread $\log_{10}(\lambda_1/\lambda_{29})$ from 7.99 to 6.92, while the numerical rank rises only from twenty to twenty-two; the Gaussian chirp produces the same effect more weakly. The chirp thus acts predominantly as an implicit preconditioner of the optimization problem, not as a source of additional information about the unknown system. That interpretation is required by the structure of the algorithm, since $\alpha(t)$ multiplies the predicted effective Hamiltonian and never enters the generation of the target trajectories, and it accounts for the pattern reported below, in which chirped modulation improves convergence and the trajectory loss but not the state-level fidelity.

One of the key results of this study, which also embodies its main aim, is presented in Figs. 6 and 7, where the dynamics obtained from the reference Hamiltonians are compared with those generated by the Hamiltonians reconstructed through the proposed learning framework and by the equivalent physical circuit. The procedure can be summarized as follows. The unknown Hamiltonian is generated by the physics-informed random sampling of Sec. II, and the QNN is trained on the full density-matrix trajectory together with the complementary strategies described above. Once training converges, the predicted Hamiltonian and the optimized dissipation coefficients define an open quantum system, which is then mapped onto a physically realizable quantum circuit. In the present study we assume that the predicted open quantum system can be represented, at least approximately, by two transmon qubits coupled through a bus resonator. Under this assumption the target Hamiltonian can be written as $H_{true} = H_{pred} - H_{extra}$, where $H_{extra}$ collects the portion of the learned Hamiltonian that cannot be represented within the two-qubit–bus-resonator architecture and is therefore treated as the mapping error. The circuit parameters are extracted from the optimized quantities $\omega_1$, $\omega_2$, $\omega_r$, $g_1$, $g_2$, $\kappa$, $\gamma_1$ and $\gamma_2$ obtained from the Hamiltonian-learning procedure, and are summarized in Tables I and II. Table I lists the transition frequencies of the two transmons ($\omega_1$ and $\omega_2$), the bus/readout resonator frequency ($\omega_r$), the dispersive shifts ($\chi_1$ and $\chi_2$), the qubit–resonator coupling rates ($g_1$ and $g_2$), the qubit–resonator detunings ($\Delta_1$ and $\Delta_2$), the effective exchange coupling ($J_{12}$), the resonator decay rate ($\kappa$) and the Purcell decay rates ($\Gamma_{P1}$ and $\Gamma_{P2}$). All entries have been recomputed self-consistently from the saved optimized parameter vector using the relations given in the table note, so that the detunings follow from the listed frequencies and every derived quantity follows from the detunings.

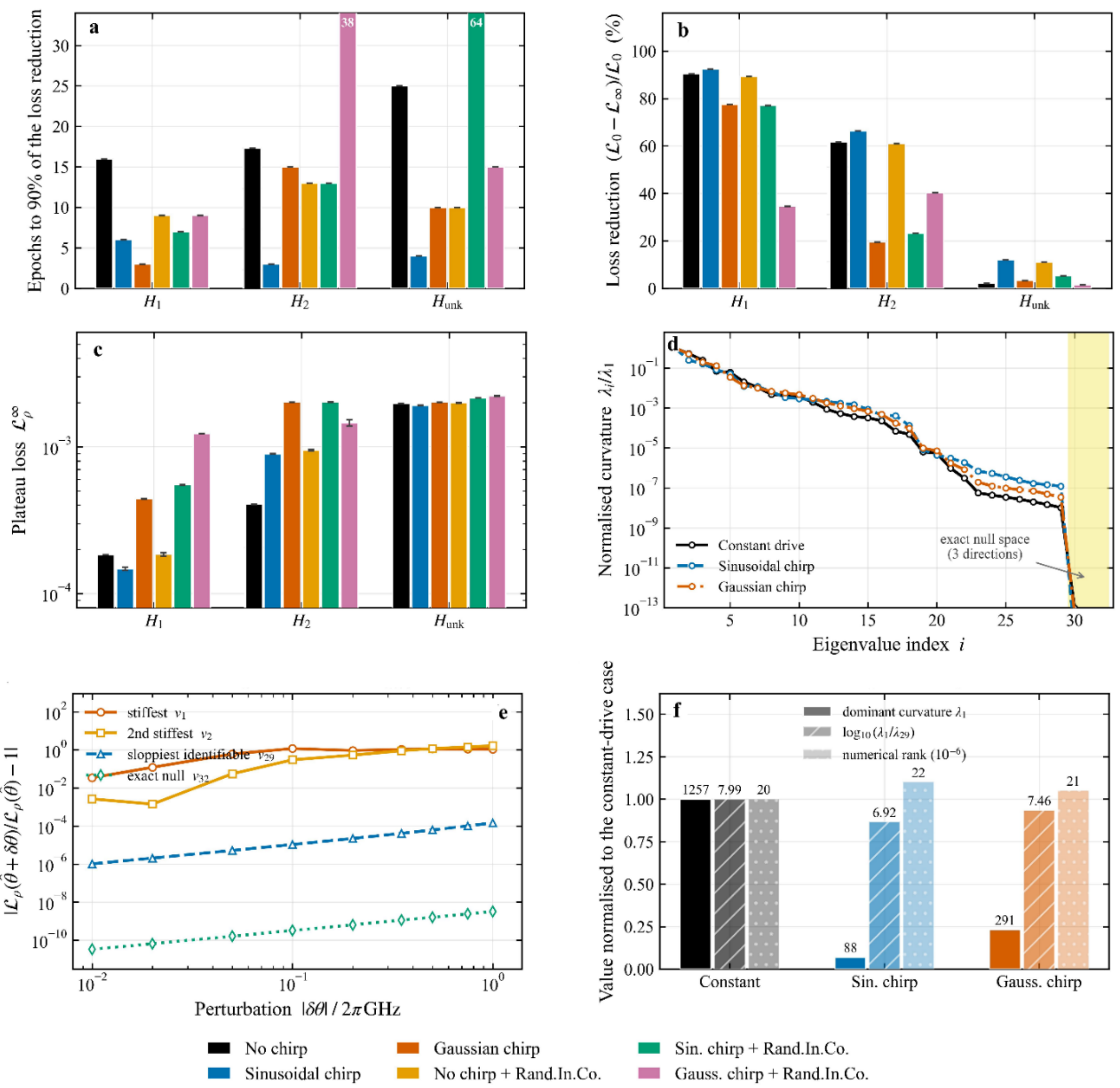

Fig. 5. Quantitative convergence behavior and structure of the optimization landscape. (a) Number of training epochs needed to realize 90% of the total loss reduction, (b) fraction of the initial loss removed by training, and (c) converged plateau loss with its standard deviation over the last twenty epochs, for the three systems and the six training strategies. (d) Normalized eigenvalue spectrum $\lambda_i/\lambda_1$ of the Gauss–Newton metric of the weighted trajectory-density loss, evaluated at the converged coefficients of the unknown system under constant, sinusoidal-chirp and Gaussian-chirp propagation; the shaded band marks the three exactly null directions, which arise because the exchange operator is the mean of the XX and YY operators. (e) Relative change of the trajectory loss when the converged coefficients are displaced along the stiffest, second-stiffest, sloppiest identifiable and an exactly null eigendirection. (f) Dominant curvature $\lambda_1$, eigenvalue spread $\log_{10}(\lambda_1/\lambda_{29})$ and numerical rank ($10^{-6}$ threshold) for the three propagation modes, each normalized to its constant-drive value; absolute values are printed above the bars.

The identified parameter set places qubit 1 comfortably in the dispersive regime, with $|\Delta_1|/g_1 = 5.4$, whereas qubit 2 sits only marginally within it, with $|\Delta_2|/g_2 = 1.8$; the dispersive approximation should therefore be regarded as quantitatively reliable for the first qubit and as indicative for the second. The same near-resonance is responsible for the asymmetric Purcell rates, $\Gamma_{P1}/2\pi = 0.98$ MHz and $\Gamma_{P2}/2\pi = 9.11$ MHz, which correspond to Purcell-limited relaxation times of about 162 ns and 17.5 ns. These values are short by the standards of state-of-the-art transmon hardware and identify the resonator linewidth and the qubit-2 detuning as the parameters that a subsequent design iteration would need to adjust; they are reported here because they are what the learning procedure returns, and not as an optimized device specification. Table II presents the derived circuit and layout parameters obtained from Table I. Using standard transmon and coplanar-waveguide relations, the total qubit capacitances ($C_{\Sigma 1}$ and $C_{\Sigma 2}$), the equivalent inter-qubit coupling capacitance ($|C_c|$), the Josephson inductances ($L_{J1}$ and $L_{J2}$), the transmon pad areas ($A_{pad,1}$ and $A_{pad,2}$), the resonator length ($L_{res}$), the equivalent resonator inductance ($L_r$) and the qubit separation ($d_{q1-q2}$) were calculated. The resulting capacitances of about 77 fF and Josephson inductances of 9.2–11.2 nH are consistent with practical superconducting transmon implementations fabricated in planar aluminum or niobium technology, and the corresponding Josephson-to-charging energy ratios, $E_J/E_C = 58$ and 71, lie well inside the transmon regime, which is an independent consistency check on the extracted parameters. The $\lambda/2$ resonator length of 15.18 mm and the qubit separation of 7.59 mm provide a physically realizable layout target for electromagnetic simulation and subsequent fabrication. Fig. 6 shows the resulting two-qubit–bus-resonator circuit. Two transmons, $Q_1$ and $Q_2$, are capacitively coupled to a common $\lambda/2$ bus/readout resonator R. The qubit transition frequencies are 5.4003 GHz and 5.9725 GHz and the resonator frequency is 6.2495 GHz, giving a loaded quality factor $Q_r = \omega_r/\kappa \approx 221$. The extracted coupling rates are $g_1/2\pi =$ 158.3 MHz and $g_2/2\pi =$ 157.2 MHz, and the effective exchange interaction is $J_{12}/2\pi = -59.56$ MHz, corresponding to an equivalent coupling capacitance of $|C_c| = 1.621$ fF. The circuit also incorporates the resonator decay and Purcell relaxation channels obtained from the learned open-system parameters. Fig. 6 therefore provides the circuit-level realization of the learned Hamiltonian and establishes the connection between the data-driven reconstruction and a realizable superconducting quantum emulator. In what follows, the Hamiltonian of this circuit is referred to as the true quantum Hamiltonian.

Table I — Optimized Quantum Hamiltonian Parameters.

| Parameter | Value | Description |
|---|---|---|
| $\omega_1/2\pi$ | 5.4003 GHz | Qubit-1 transition frequency |
| $\omega_2/2\pi$ | 5.9725 GHz | Qubit-2 transition frequency |
| $\omega_r/2\pi$ | 6.2495 GHz | Bus/readout resonator frequency |
| $\chi_1$ | −6.71 MHz | Dispersive shift of qubit-1 |
| $\chi_2$ | −42.28 MHz | Dispersive shift of qubit-2 |
| $g_1$ | 158.3 MHz | Qubit-1–resonator coupling |
| $g_2$ | 157.2 MHz | Qubit-2–resonator coupling |
| $\Delta_1$ | −849.3 MHz | Qubit-1 detuning |
| $\Delta_2$ | −277.1 MHz | Qubit-2 detuning |
| $J_{12}$ | −59.56 MHz | Exchange coupling |
| $\kappa$ | 28.30 MHz | Resonator decay rate |
| $\Gamma_{p1}$ | 0.98 MHz | Purcell decay rate of qubit-1 ($\Gamma_{p1}/2\pi$) |
| $\Gamma_{p2}$ | 9.11 MHz | Purcell decay rate of qubit-2 ($\Gamma_{p2}/2\pi$) |
| $\alpha_1$ | ~−250 MHz | Qubit-1 anharmonicity |
| $\alpha_2$ | ~−250 MHz | Qubit-2 anharmonicity |

Table II: Derived superconducting circuit and layout parameters extracted from the optimized Hamiltonian.

| Parameter | Value | Description |
|---|---|---|
| $C_{\Sigma_1}$ | 77.3 fF | Total capacitance of qubit-1 |
| $C_{\Sigma_2}$ | 77.3 fF | Total capacitance of qubit-2 |
| $\lvert C_c\rvert^*$ | 1.621 fF | Inter-qubit coupling capacitance |
| $L_{J1}$ | 11.21 nH | Josephson inductance of qubit-1 |
| $L_{J2}$ | 9.17 nH | Josephson inductance of qubit-2 |
| $A_{pad,1}$ | 436 µm² | Estimated transmon pad area |
| $A_{pad,2}$ | 436 µm² | Estimated transmon pad area |
| $L_{res}$ | 15.18 mm | λ/2 CPW resonator length |
| $d_{q1-q2}$ | 7.59 mm | Estimated qubit separation |
| $L_r$ | 0.811 nH | Equivalent resonator inductance |

Notes on Tables I and II. All values are recomputed self-consistently from the optimized parameter vector returned by the learning procedure. The identified Hamiltonian parameters are angular frequencies in rad $s^{-1}$; the GHz and MHz values shown are obtained by dividing by 2π. The detunings follow directly from the listed frequencies, $\Delta_i = \omega_i - \omega_r$. The dispersive shifts use the transmon expression $\chi_i = g_i^2\alpha_i / [\Delta_i(\Delta_i + \alpha_i)]$ with $\alpha_i/2\pi = -250$ MHz, and the exchange coupling uses $J_{12} = (g_1g_2/2)(1/\Delta_1 + 1/\Delta_2)$. The Purcell rates are estimated from $\Gamma_{P,i} = (g_i/\Delta_i)^2\kappa$ and are quoted here, as everywhere in this work, divided by 2π. The λ/2 resonator length uses $\varepsilon_{eff} = 2.5$, and the qubit separation is taken as Lres/2. The total qubit capacitance follows from $C_\Sigma = e^2/(2E_C)$ with $E_C = \hbar|\alpha|$, the Josephson inductance from $L_J = \Phi_0^2/(4\pi^2E_J)$ with $E_J = \hbar\omega^2/(8|\alpha|)$, the pad area from the parallel-plate estimate $A_{pad} = C_\Sigma t_{ox}/\varepsilon_0$ with $t_{ox} = 50$ nm, and the equivalent resonator elements from $C_r = \pi/(2Z_0\omega_r)$ and $L_r = 1/(\omega_r^2C_r)$ with $Z_0 = 50\ \Omega$. The magnitude $|C_c|$ is reported because the negative sign of the exchange coupling arises from the chosen capacitance-matrix convention and does not represent a physically negative capacitor. The capacitance, pad-area and layout estimate rest on the stated lumped-element and parallel-plate approximations and are intended as design targets for electromagnetic simulation rather than as final fabrication values.

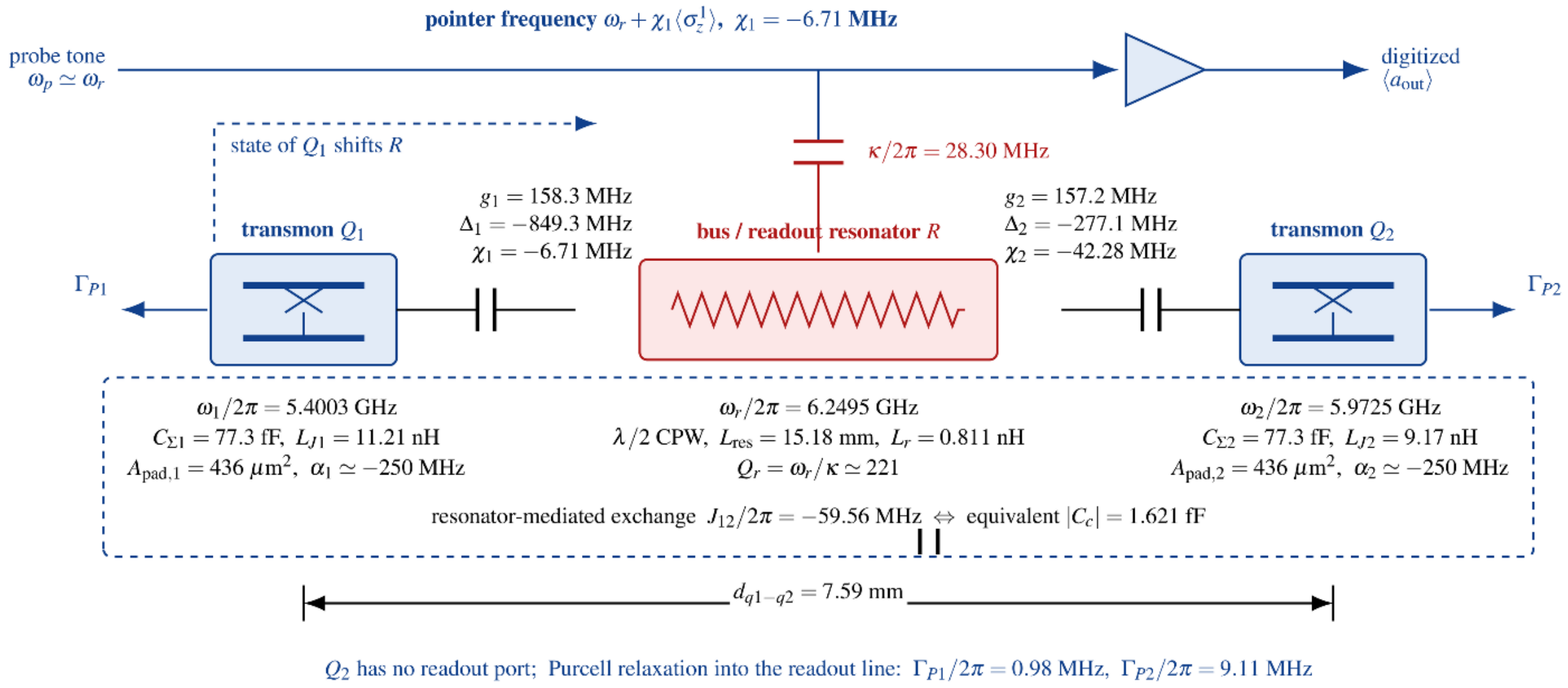

Fig. 6. Physical two-qubit–bus-resonator quantum circuit reconstructed from the optimized Hamiltonian and dissipation parameters of Tables I and II. Two transmon qubits $Q_1$ and $Q_2$ are capacitively coupled to a common λ/2 coplanar-waveguide bus/readout resonator R, which is probed through a feedline; the state of $Q_1$ shifts the resonator pointer frequency by $\chi_1$. All values shown are those returned by the learning procedure, recomputed self-consistently as described in the note to Tables I and II. The dispersive, exchange, Purcell and transmon relations used here are the standard circuit-QED and transmon expressions [84, 85].

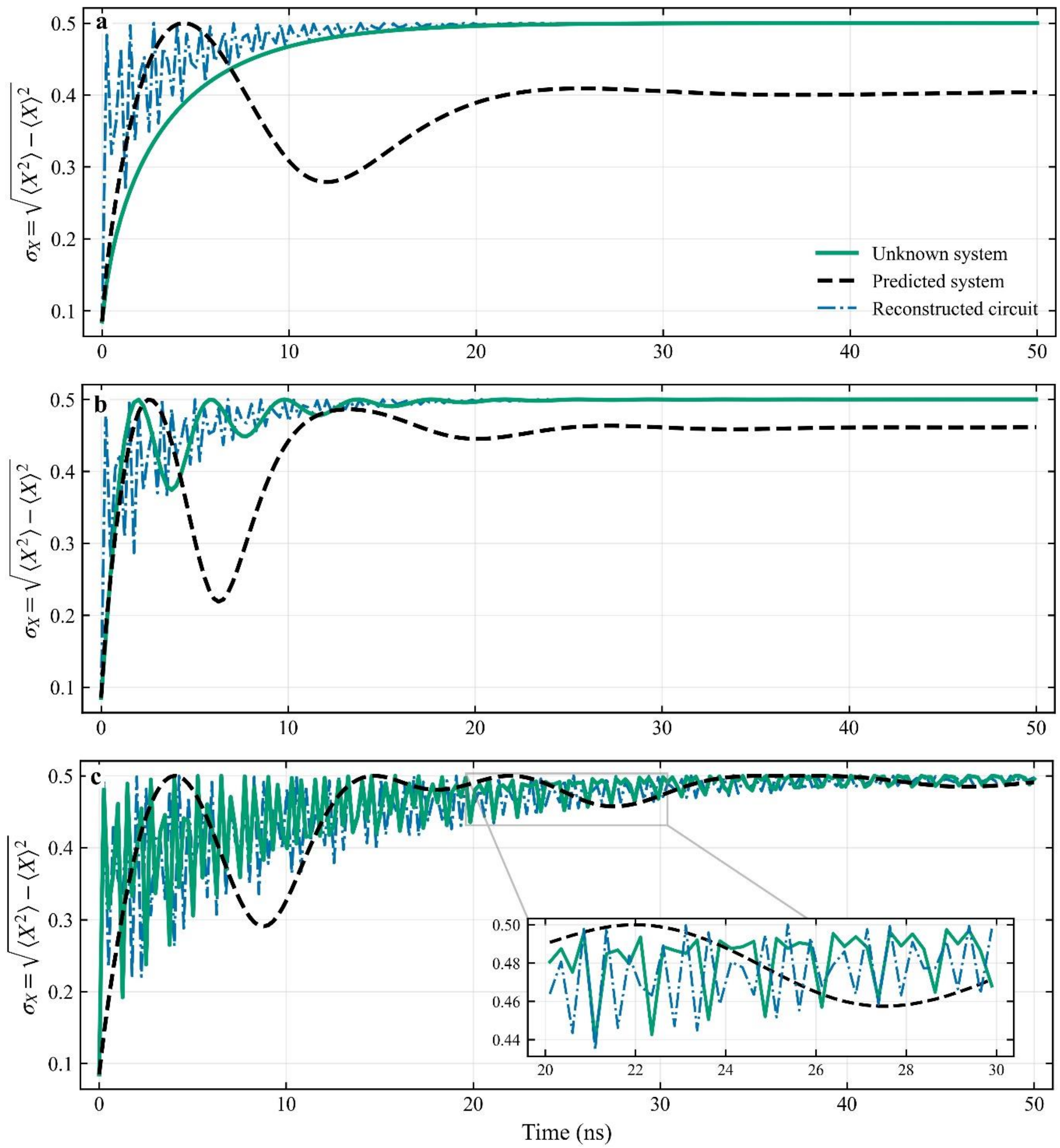

Fig. 7. Comparison of the resonator quadrature fluctuation $\sigma_X = [\langle X^2\rangle - \langle X\rangle^2]$^(1/2) obtained from the reference system (green), from the Hamiltonian predicted by the QNN (black dashed) and from the reconstructed two-qubit–bus-resonator circuit (blue dash-dotted): (a) unknown Hamiltonian equal to H1, (b) unknown Hamiltonian equal to $H_2$, and (c) the genuinely unknown Hamiltonian ($H_{unk}$) defined by thirty-two randomly sampled coefficients. The inset in (c) magnifies the interval 20–30 ns.

Fig. 7a corresponds to the single-qubit Hamiltonian $H_1$, Fig. 7b to the coupled-qubit system $H_2$, and Fig. 7c to the genuinely unknown Hamiltonian. In each panel the solid green curve is the reference response of the system under test ($H_1$ in Fig. 7a, $H_2$ in Fig. 7b, and $H_{unk}$ in Fig. 7c), the black dashed curve is the response obtained from the Hamiltonian predicted by the QNN, and the blue dash-dotted curve is the response of the reconstructed two-qubit–bus-resonator circuit. As described above, the procedure first learns the system dynamics from the full density-matrix trajectory and then converts the optimized parameters into a physical circuit intended to reproduce the behaviour of the reference system. It should be emphasized that the predicted Hamiltonian is always constructed within the 32-operator basis of Eq. 3. Consequently, even for a system as simple as $H_1$, the reconstruction is performed in a parameter space far larger than the one physically required, and the resulting over-parameterization can produce a substantial deviation in individual coefficients or in an observable such as the quadrature fluctuation of Fig. 7a. This deviation must be distinguished from the trajectory-reconstruction error discussed in Figs. 2–5, because the two quantities probe different aspects of the problem: the former reflects the accuracy and uniqueness of the Hamiltonian-parameter estimate, whereas the latter measures how closely the predicted system reproduces the observed dynamics under the training objective. The important observation from Fig. 7 is that the reconstructed circuit follows the principal dynamical behavior of the reference system, including the initial transient and the approach to the steady state, which demonstrates the feasibility of building a physical quantum twin from the learned Hamiltonian and the optimized dissipation parameters. Nevertheless, the agreement degrades as the complexity of the target Hamiltonian increases, and the residual offset visible in the steady-state level of Fig. 7a and 7b is a direct manifestation of the identifiability limits quantified in Fig. 5 and analysed further in Figs. 8–10. This behavior highlights the intrinsic difficulty of mapping a higher-dimensional learned Hamiltonian onto a constrained physical architecture on the basis of finite dynamical data.

As a complementary task, several quantum-state validation metrics are now used to assess the performance of the proposed QNN more rigorously. Since no single metric can completely characterize the agreement between two quantum states, two well-established quantities from quantum information theory are considered alongside the trajectory-density loss: the quantum-state fidelity and the trace distance. Each captures a different physical aspect of the reconstructed density matrix, namely the state overlap and the operational distinguishability. Similar validation methodologies are widely adopted in quantum tomography, quantum control and Hamiltonian-learning studies [23, 48–53]. The reconstruction of an unknown Hamiltonian cannot be assessed reliably from the training loss alone, because a small density-matrix reconstruction error does not by itself guarantee that the reconstructed system reproduces the physically relevant properties of the unknown dynamics. Fidelity quantifies the overlap between the density matrices of the unknown and reconstructed systems, with $F = 1$ for identical states, whereas trace distance quantifies their distinguishability, with $D = 0$ for identical states; the relevant definitions are given in Appendix A and Appendix B. The results in Table III show that the reconstruction accuracy depends on both the underlying Hamiltonian and the excitation or initialization strategy. For $H_1$ under sinusoidal-chirp modulation the reconstructed system reaches $F = 0.847609$ with $D = 0.305049$. When randomized initial conditions are introduced the fidelity rises substantially to $F = 0.929136$ and the trace distance falls to $D = 0.124056$, a clear improvement in state-level agreement. A similar trend is observed for $H_2$: sinusoidal-chirp modulation yields $F = 0.889151$ and $D = 0.258429$, whereas randomized initial conditions give $F = 0.884629$ and $D = 0.193076$, so that the fidelity is essentially unchanged while the trace distance improves by a quarter. For the genuinely unknown Hamiltonian the sinusoidal chirp gives $F = 0.708173$ and $D = 0.400712$, the largest discrepancy among the investigated cases, and randomized initial conditions improve the reconstruction to $F = 0.787372$ and $D = 0.316218$. Consequently, the randomized initial conditions improve the agreement between the reconstructed and reference quantum states, particularly for the more complex unknown Hamiltonian. This trend differs from the one observed for the trajectory-density loss in Figs. 2–5, where sinusoidal-chirp modulation generally produces a lower training loss than randomized initialization. The apparent discrepancy is not a contradiction but a direct consequence of the analysis of Sec. II and Fig. 5: the chirp acts on the conditioning of the optimization and cannot add information about the unknown system, whereas randomizing the initial state changes the data themselves and therefore the region of Hilbert space over which the model is constrained. A lower training loss consequently does not imply a higher final-state fidelity or a lower trace distance, and the two families of metrics must be reported jointly. How far these state-level scores are from the best that the present formulation can achieve is quantified in Fig. 8.

Table III. Quantum-state reconstruction metrics for the investigated quantum systems.

| Hamiltonian | Condition | Fidelity | Trace Distance |
|---|---|---|---|
| **$H_1$** | With Sin. Chirp | 0.8476 | 0.3050 |
| **$H_1$** | Rand.In.Co. | 0.9291 | 0.1240 |
| **$H_2$** | With Sin. Chirp | 0.8891 | 0.2584 |
| **$H_2$** | Rand.In.Co. | 0.8846 | 0.1930 |
| **Un Known** | With Sin. Chirp | 0.7081 | 0.4007 |
| **Un Known** | Rand.In.Co. | 0.7873 | 0.3162 |

**Note:** fidelity and trace distance quantify the agreement between the reconstructed and the unknown density matrix at the end of the 50 ns evolution window, evaluated for the best-matching realization encountered during training. The complete set of runs, together with the trajectory-level metrics and the attainability bounds, is collected in Table IV.

The fidelities of Table III range from 0.71 to 0.93, which invites the question of how much of the residual gap is an artefact of the learning procedure and how much is imposed by the formulation itself. Fig. 8 answers this question quantitatively. As established in Sec. II, the differentiable propagator of Eq. 5 renormalizes the state after every step and therefore returns a rank-one density matrix; the predicted von Neumann entropy is numerically zero in every one of the eighteen runs analyzed here. For a pure prediction the Uhlmann fidelity reduces to the expectation value $\langle\psi|\rho_{unk}|\psi\rangle$, so that $F \leq \lambda_{max}$, and the Fuchs–van de Graaf inequality [52, 63] then gives $D \geq 1 - F \geq 1 - \lambda_{max}$. Fig. 8a and Fig. 8b show that every measured pair (F, D) respects these bounds, and that the achieved fidelities cluster close to the diagonal $F = \lambda_{max}$. Fig. 8c reports the resulting normalized fidelity $F/\lambda_{max}$, which measures the fraction of the attainable overlap that the learning procedure actually reaches. For the single-qubit benchmark this ratio is 0.97–1.00 for every strategy: the identified model is essentially as close to the target as any pure state can be, and the 7% fidelity deficit of the randomized-initialization run, F = 0.9291, is due almost entirely to the mixedness of the target state, whose largest eigenvalue has fallen to 0.936 by the end of the evolution window. For $H_2$ the ratio is 0.77–0.98 and for the unknown system 0.61–0.94, so that the genuine learning deficit is between two and four times smaller than the raw fidelity numbers suggest. Two conclusions emerge from these results. First, a substantial fraction of the residual reported in Table III appears to arise from structural limitations rather than from the learning algorithm itself. Consequently, the most effective approach for improving the state-level metrics would be to replace the no-jump propagator with a differentiable density-matrix solver capable of representing mixed states, albeit at the expense of increased memory requirements and computational cost. Second, the normalized fidelity, ($F/\lambda_{max}$), provides a more appropriate figure of merit when comparing runs with target states of different purities. Therefore, it is suggested to reporting $F/\lambda_{max}$ alongside the raw fidelity in trajectory-learning studies of open quantum systems. Fig. 8d addresses the complementarity of the metrics directly, using the complete record of best-so-far updates from one training run. The trajectory loss and fidelity are strongly anticorrelated overall, with a Spearman rank correlation [83] of (-0.97), indicating that minimizing the trajectory loss is a reliable proxy for improving state-level agreement. However, this correspondence does not hold consistently at the individual-update level: among the 127 updates that reduced the trajectory loss, 59 were accompanied by a decrease in the fidelity of the corresponding state. This result quantitatively supports the qualitative observation above and demonstrates that, despite their strong global correlation, the two metrics capture distinct aspects of performance and therefore cannot be used interchangeably.

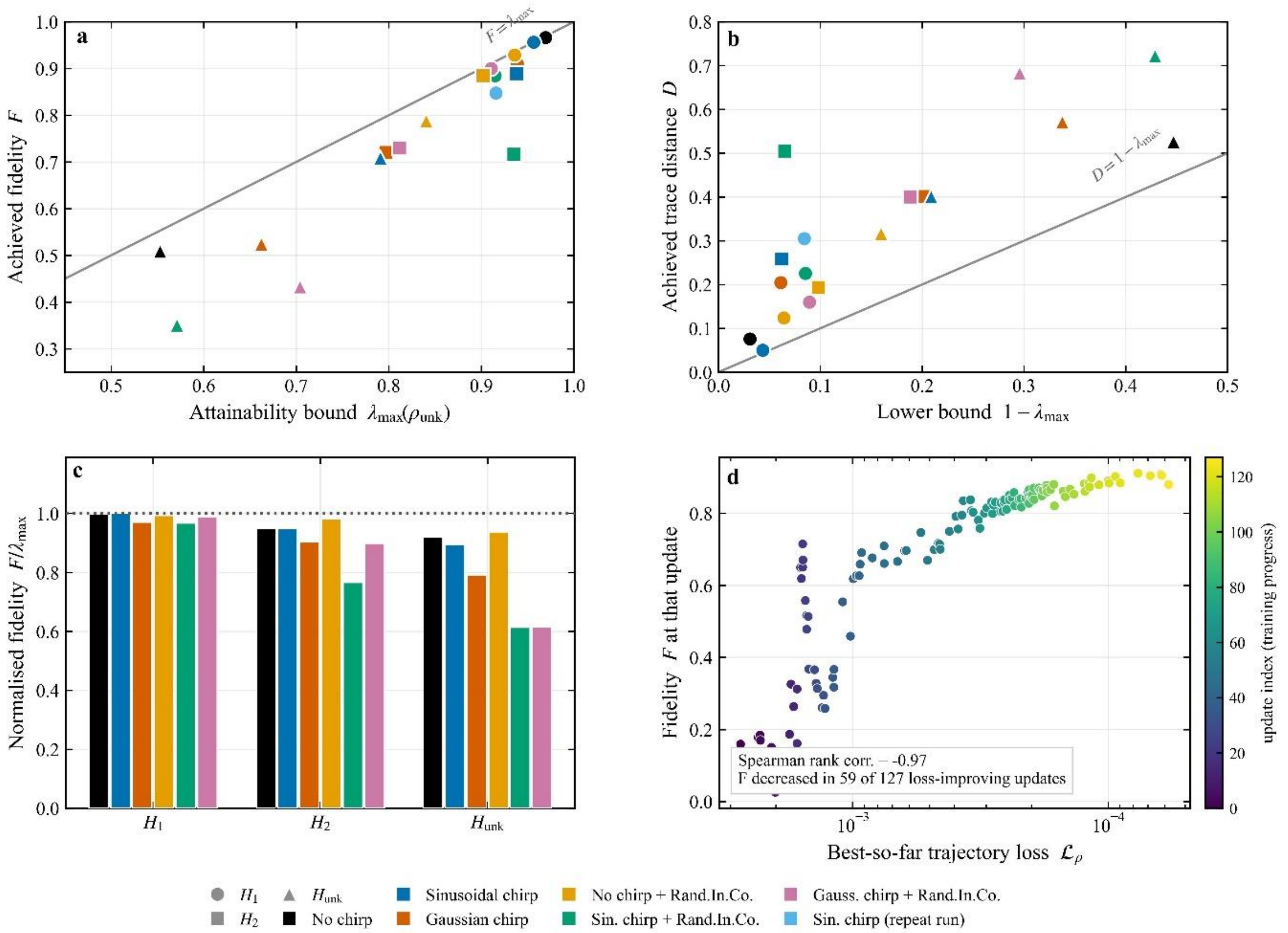

Fig. 8. Attainability of the state-level metrics and complementarity of the loss and fidelity criteria. (a) Achieved fidelity versus the purity bound $\lambda_{\max}$ for all eighteen training runs; marker shape denotes the system and color the training strategy, and the grey line is the attainability limit $F = \lambda_{\max}$. (b) Achieved trace distance versus its lower bound $1 - \lambda_{\max}$. (c) Normalized fidelity $F/\lambda_{\max}$, i.e. the fraction of the attainable overlap reached by each run; the dotted line marks the limit. (d) Fidelity plotted against the best-so-far trajectory loss along a complete training run, colored by update index.

Fig. 5 and Fig. 8 characterize the optimization and the state-level agreement. Fig. 9 turns to the operator itself and asks how well the learned Hamiltonian reproduces the unknown one, which is the question that matters if the identified model is to serve as a digital twin rather than as a trajectory fit. Fig. 9a–Fig. 9c compare the magnitudes of the unknown Hamiltonian, the learned Hamiltonian, and their difference, respectively, for the 32-coefficient unknown system trained using a sinusoidal chirp. Both operators are dominated by their diagonal, as expected from the physics-informed weighting of the basis, and both populate the same off-diagonal blocks, so the learned operator is structurally plausible; the residual, however, is of the same order as the operator itself. Fig. 9d provides a quantitative comparison at the level of the energy spectrum. The learned Hamiltonian covers a similar overall energy range, from −1.87 to 2.18 GHz, compared with −1.73 to 2.21 GHz for the target Hamiltonian. However, the internal level ordering differs substantially: the learned spectrum increases sharply between the fourth and sixth levels before flattening at higher levels. Consequently, the individual level spacings, and therefore the transition frequencies that would be observed spectroscopically, are not accurately reproduced.

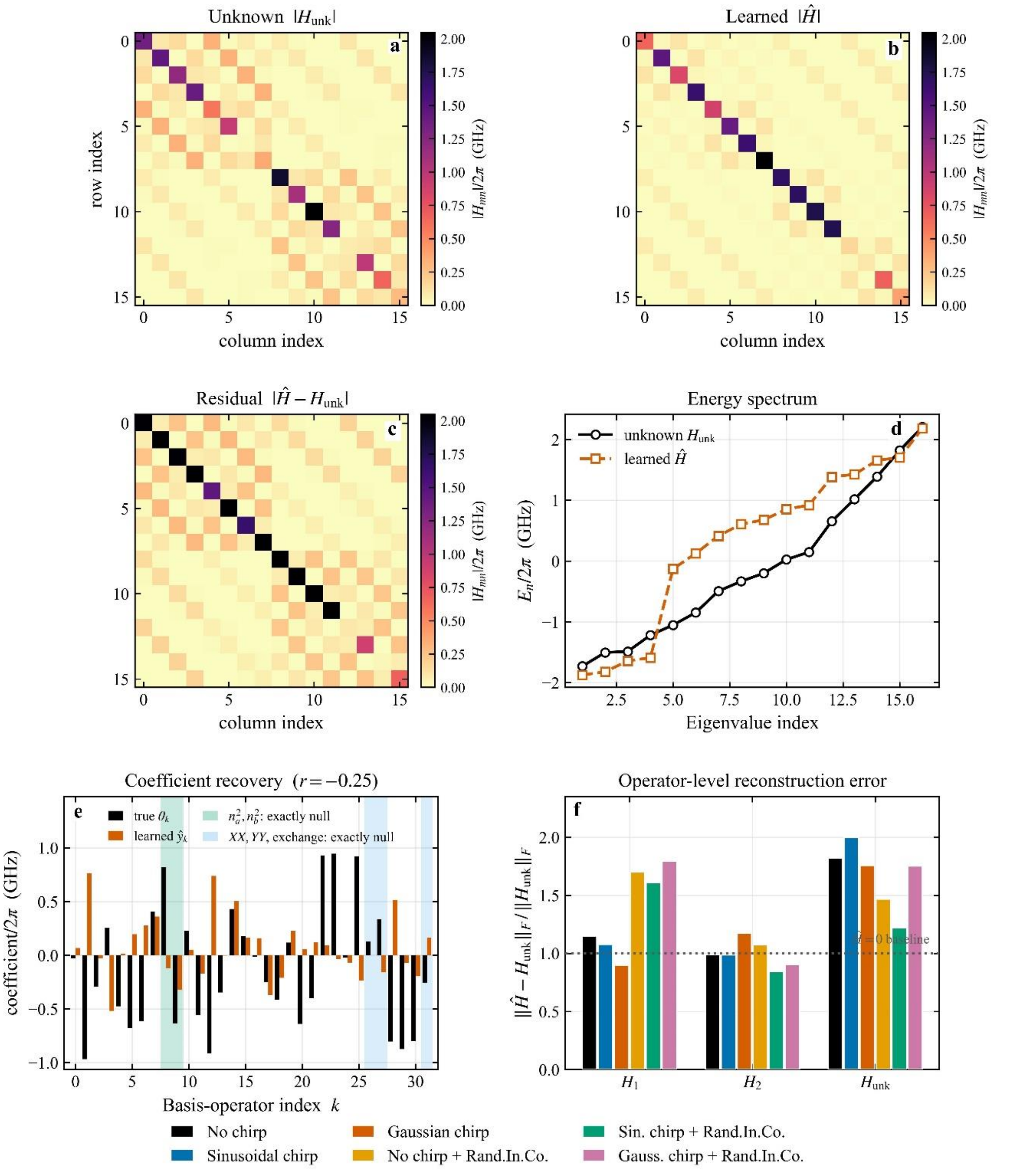

Fig. 9. Operator-level reconstruction and parameter recovery for the 32-coefficient unknown system trained with sinusoidal chirp. (a) Modulus of the unknown Hamiltonian, (b) of the learned Hamiltonian and (c) of their difference, all in units of 2π GHz and on a common color scale. (d) Energy spectra of the unknown and learned Hamiltonians. (e) True and learned coefficients over the 32-operator basis; the shaded bands mark the operator groups that span the exact null space of the basis and are therefore individually unrecoverable in principle. (f) Relative operator error $\|\hat{H} - H_{unk}\|_F / \|H_{unk}\|_F$ for every system and strategy; the dotted line is the error obtained by returning the zero operator.

Fig. 9e compares the learned coefficients with the true ones directly. The correlation between the two vectors is −0.25, and across the six unknown-system runs it ranges only from −0.25 to +0.17; the relative coefficient error is of order unity in every case. The shaded bands mark the three groups involved in the exact null directions of the basis identified in Sec. II, within which no learning algorithm could recover the individual coefficients even from noiseless, infinitely long data. Fig. 9f collects the relative operator error $||\hat{H} - H_{unk}||_F / ||H_{unk}||_F$ [64, 81, 82] for all runs. The values lie between 0.84 and 2.00, and the dotted line marks the error that would be obtained by returning the zero operator. The learned Hamiltonians are therefore not close to the unknown ones in operator norm, even where the trajectory loss is small and the state fidelity is close to its attainable bound. This is the central negative result of the study, and it is a consequence of the analysis in Fig. 5d–Fig. 5f rather than of a failure of training: with three exactly null and roughly ten numerically flat directions, the trajectory data constrain only a low-dimensional projection of the 32-dimensional coefficient vector, and any of a large family of Hamiltonians reproduces the observed dynamics almost equally well. The proposed framework should accordingly be understood as a method for constructing a behavioral emulator of an unknown open quantum system – a model that reproduces its measurable dynamics within a stated operating regime – and not as a method for uniquely determining the microscopic Hamiltonian. Recovering the coefficients themselves would require breaking these degeneracies, for example by removing the redundant operators from the basis, by probing the system from a set of initial states that span more of the Hilbert space, or by adding physically motivated sparsity or hierarchy priors to the coefficient vector.

The final validation concerns the quantity that the framework is ultimately intended to deliver, namely the agreement between the predicted and the measured dynamics over the whole evolution rather than at a single instant. In line with, Fig. 10 reports the time-resolved fidelity and trace distance for the three systems. Fig. 10a and Fig. 10b evaluate the surrogate model that is actually trained, that is the chirped no-jump propagator of Eq. 5, against the target Lindblad trajectory; the shaded bands span twelve independent draws of the chirp parameters from the ranges specified in Sec. II. For the single-qubit benchmark the fidelity settles at 0.85 on average and tracks its purity bound, shown dotted, to within 0.07 over most of the window. For $H_2$ and for the unknown system the mean fidelities are 0.66 and 0.61, and the bands are wide, which shows that the realized chirp parameters have a strong influence on the achieved trajectory agreement; this run-to-run sensitivity is an additional reason for reporting distributions rather than single numbers, and it is consistent with the variability observed between nominally identical runs in Table IV. Fig. 10c displays the purity of the target and of the prediction. The target purity decays from unity to 0.84, 0.88 and 0.65 for $H_1$, $H_2$ and the unknown system respectively, whereas the prediction remains at unity throughout, which is the rank-one constraint of Fig. 8 seen in the time domain and is the dominant contribution to the residual trace distance at long times. Fig. 10d applies the most demanding test available: the learned Hamiltonian is removed from the surrogate propagator and re-inserted into the full Lindblad equation, so that it is used exactly as a physical model would be. For the single-qubit benchmark the trajectory-averaged fidelity is 0.94 and the agreement persists over the whole window, which demonstrates that for a system whose effective dimensionality matches the information content of the data the learned operator is a genuine dynamical model and not merely a fit. For $H_2$ and the unknown system the trajectory-averaged fidelities fall to 0.57 and 0.58. The transfer from the training surrogate to a true open-system propagator is therefore only partial for the multi-mode cases, which is consistent with the operator-level errors of Fig. 9 and with the fact that the chirped propagator and the Lindblad generator are not equivalent dynamical maps. This delineates the present operating envelope of the framework: reliable behavioral emulation for systems whose effective dimensionality is small compared with the size of the operator basis, and progressively weaker transfer as that gap widens.

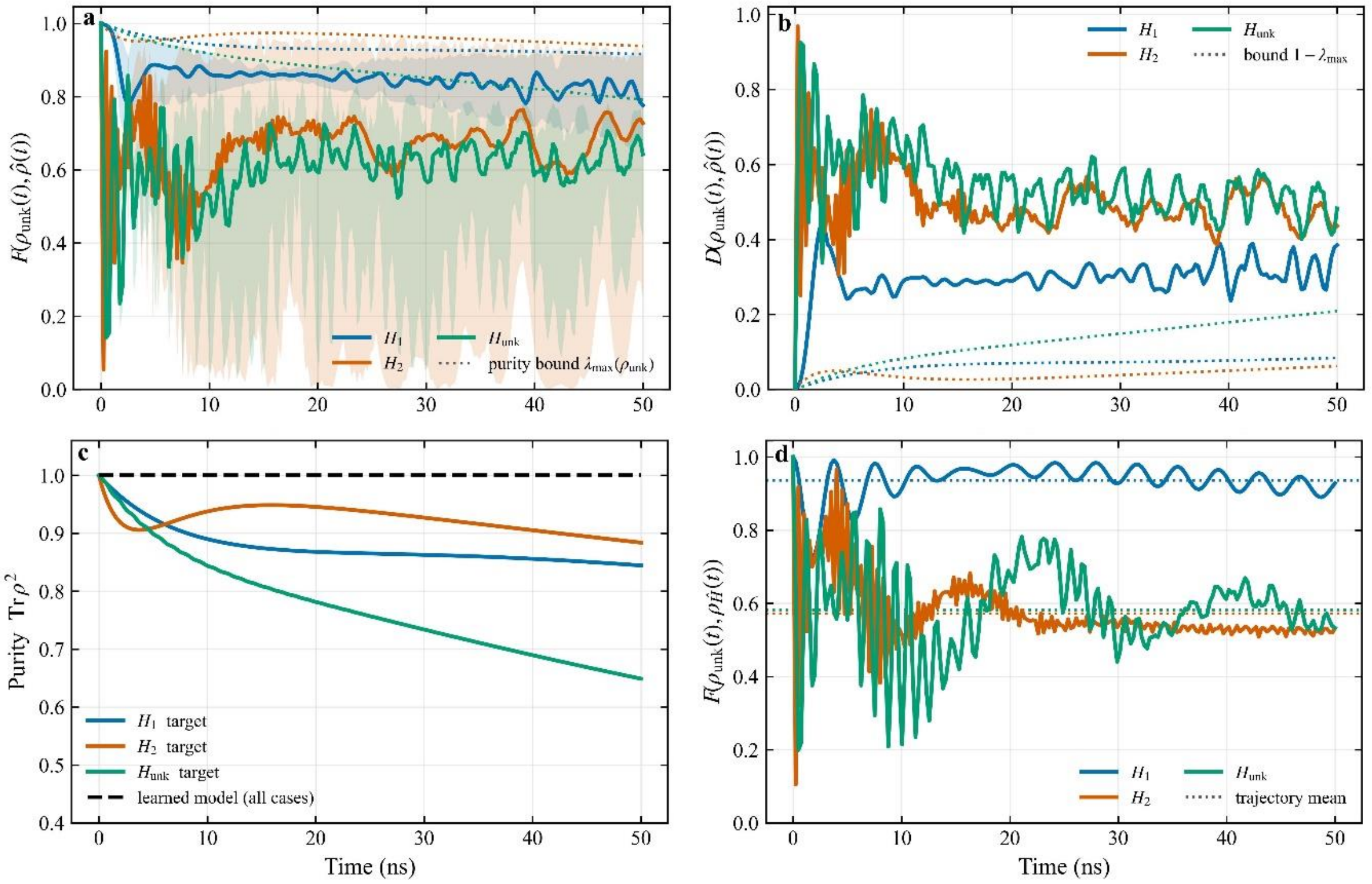

Fig. 10. Trajectory-level validation of the learned models. (a) Fidelity and (b) trace distance between the target Lindblad trajectory and the trajectory produced by the trained surrogate propagator of Eq. 5, for $H_1$, $H_2$ and the unknown system; the shaded bands span twelve independent draws of the chirp parameters and the dotted curves are the corresponding purity bounds $\lambda_{max}(\rho_{unk})$ and $1 - \lambda_{max}$. (c) Purity of the target trajectories compared with that of the prediction, which is unity at all times because the surrogate propagator returns a rank-one state. (d) Emulator test: the learned Hamiltonian is re-propagated with the full Lindblad equation under the identified dissipation rates and compared with the target; dotted lines mark the trajectory-averaged fidelities of 0.94, 0.57 and 0.58.

The analyses presented above clarify the capabilities and limitations of the proposed framework. It is therefore important to explicitly define the scope of the present study and identify the aspects that remain beyond its current reach. First, the differentiable propagator is a no-jump, norm-renormalized quantum-trajectory solver rather than a full density-matrix solver, so the predicted state is always pure. This bounds the achievable fidelity by the purity of the target state and is responsible for a large part of the residual reported in Table III; extending the propagator to a differentiable Lindblad or vectorized-superoperator solver is the most direct improvement available. Second, the 32-operator basis is rank-deficient by construction, and the trajectory data constrain only about twenty of the twenty-nine remaining directions at the $10^{-6}$ level. Individual Hamiltonian coefficients are therefore not identifiable, and the framework delivers a behavioral emulator rather than a unique microscopic model. Third, the control vector supplied to the network is drawn independently of the system that produced each trajectory, so the mapping from control settings to coefficients cannot generalize across systems; the results reported here concern the identification of individual systems and the ensemble-average fit, and conditioning the network on measured features of the trajectory itself is a natural next step. Fourth, the state-level metrics of Table III are evaluated for the best-matching realization encountered during training, and the spread between nominally identical runs visible in Table IV shows that single-run numbers carry appreciable variance; larger ensembles of runs would be needed for statistically firm comparisons between strategies. Fifth, the injected measurement noise contributes a loss floor more than three orders of magnitude below the achieved residual, so the present results do not constitute a test of robustness against realistic measurement noise. Finally, both bosonic modes are truncated to two levels, which is what makes the Kerr operators degenerate with the number operators and which limits the applicability of the identified model to the low-excitation regime. None of these limitations affects the central positive results – the demonstrated convergence behavior, the mechanism by which chirped modulation acts, the attainability analysis, and the successful projection onto a realizable circuit – but all of them bound the interpretation of the reconstructed Hamiltonian.

Table IV. Complete set of training runs. For each system and strategy the table lists the fraction of the initial trajectory-density loss removed by training, the converged plateau loss, the best single-realization loss, the corresponding fidelity and trace distance, the attainability bound $\lambda_{max}(\rho_{unk})$, the normalized fidelity $F/\lambda_{max}$, the relative operator error and the trajectory-averaged fidelity obtained when the learned Hamiltonian is re-propagated with the full Lindblad equation.

| System | Training strategy | Loss red. (%) | Plateau $\mathscr{L}_P$ ($\times10^{-4}$) | F | $\lambda_{max}$ | $F/\lambda_{max}$ | D | Rel. H error | $\bar{F}$ (Lindblad) |
|---|---|---|---|---|---|---|---|---|---|
| $H_1$ | No chirp | 90.4 | 1.85 | 0.967 | 0.969 | 0.997 | 0.076 | 1.15 | 0.95 |
| $H_1$ | Sinusoidal chirp | 92.5 | 1.48 | 0.956 | 0.957 | 1.000 | 0.050 | 1.08 | 0.66 |
| $H_1$ | Gaussian chirp | 77.5 | 4.42 | 0.910 | 0.939 | 0.970 | 0.205 | 0.90 | 0.92 |
| $H_1$ | No chirp + Rand.In.Co. | 89.5 | 1.85 | 0.929 | 0.936 | 0.993 | 0.124 | 1.70 | 0.80 |
| $H_1$ | Sin. chirp + Rand.In.Co. | 77.1 | 5.50 | 0.885 | 0.915 | 0.967 | 0.226 | 1.61 | 0.91 |
| $H_1$ | Gauss. chirp + Rand.In.Co. | 34.6 | 12.30 | 0.900 | 0.911 | 0.989 | 0.160 | 1.79 | 0.48 |
| $H_2$ | No chirp | 61.7 | 4.06 | 0.625 | 0.725 | 0.948 | 0.185 | 0.991 | 0.45 |
| $H_2$ | Sinusoidal chirp | 66.4 | 8.96 | 0.889 | 0.938 | 0.948 | 0.258 | 0.99 | 0.58 |
| $H_2$ | Gaussian chirp | 19.3 | 20.13 | 0.720 | 0.797 | 0.904 | 0.401 | 1.17 | 0.49 |
| $H_2$ | No chirp + Rand.In.Co. | 61.4 | 9.47 | 0.885 | 0.902 | 0.981 | 0.193 | 1.07 | 0.80 |
| $H_2$ | Sin. chirp + Rand.In.Co. | 23.3 | 20.15 | 0.716 | 0.935 | 0.766 | 0.504 | 0.84 | 0.45 |
| $H_2$ | Gauss. chirp + Rand.In.Co. | 46.5 | 14.57 | 0.730 | 0.812 | 0.899 | 0.400 | 0.90 | 0.67 |
| $H_{unk}$ | No chirp | 2.1 | 19.67 | 0.509 | 0.553 | 0.920 | 0.526 | 1.82 | 0.52 |
| $H_{unk}$ | Sinusoidal chirp | 12.5 | 19.10 | 0.708 | 0.791 | 0.895 | 0.401 | 2.00 | 0.58 |
| $H_{unk}$ | Gaussian chirp | 3.3 | 20.09 | 0.524 | 0.662 | 0.791 | 0.571 | 1.76 | 0.59 |
| $H_{unk}$ | No chirp + Rand.In.Co. | 11.2 | 19.94 | 0.787 | 0.840 | 0.937 | 0.316 | 1.47 | 0.64 |
| $H_{unk}$ | Sin. chirp + Rand.In.Co. | 5.3 | 21.55 | 0.351 | 0.571 | 0.614 | 0.722 | 1.22 | 0.37 |
| $H_{unk}$ | Gauss. chirp + Rand.In.Co. | 1.2 | 22.18 | 0.433 | 0.704 | 0.614 | 0.683 | 1.75 | 0.46 |

Loss red.: fraction of the initial trajectory-density loss removed by training. Plateau $\mathscr{L}_P$: converged loss level. λ max: largest eigenvalue of the target density matrix, i.e. the attainable fidelity. Rel. H error = $\|H_{hat} - H_{unk}\|_F / \|H_{unk}\|_F$. $\bar{F}$ (Lindblad): trajectory-averaged fidelity obtained when $\hat{H}$ is re-propagated with the full Lindblad equation.

## Conclusions

This study introduced a unified QNN-based framework for the identification and emulation of unknown open quantum systems using full density-matrix trajectory learning. Unlike approaches based on final-state information or on individual observables, the framework exploits the complete temporal evolution of the density matrix under Lindblad dynamics, so that both coherent and dissipative features of the unknown system contribute to the identification. A physics-informed stochastic dataset was constructed by generating diverse Hamiltonians within a structured 32-operator basis together with resonator and qubit dissipation parameters, and measurement noise, chirped modulation and randomized initial quantum states were incorporated to increase the diversity and realism of the training trajectories. The QNN generates thirty-two Hamiltonian coefficients, which are assembled into a Hermitian operator and propagated through a differentiable open-system evolution, and the resulting trajectory is matched to the measured one at every sampled time. The numerical results show that the framework learns progressively more complex systems, with the trajectory-density loss increasing as the effective dimensionality of the Hamiltonian grows. Sinusoidal-chirp modulation provided the most favorable trajectory-learning performance, shortening convergence from sixteen epochs to six and lowering the plateau loss for the single-qubit benchmark, whereas the combined chirp and randomized-initialization strategy converged quickly but settled at a higher loss. Randomized initial conditions instead improved the state-level reconstruction, most clearly for the unknown Hamiltonian, where the fidelity rose from 0.7081 to 0.7873 and the trace distance fell from 0.4007 to 0.3162. A direct analysis of the optimization landscape explains this division of labor: chirped modulation reduces the dominant curvature of the trajectory loss by more than an order of magnitude and compresses its eigenvalue spread, acting as an implicit preconditioner, while leaving the information content of the data unchanged, since the modulation is applied to the predicted evolution and not to the system that generates the measurements. Beyond these strategy-level results, the study establishes the limits within which such reconstructions should be interpreted. Because the differentiable propagator returns a rank-one state, the achievable fidelity is bounded by the largest eigenvalue of the target density matrix; the measured fidelities reach between 0.61 and 1.00 of that bound and exceed 0.90 in twelve of eighteen runs, so most of the apparent residual is structural rather than algorithmic. Conversely, the operator-level error remains of order unity in every run, and a sensitivity analysis identifies three exactly null and about ten numerically flat directions in the coefficient space, showing that the measured trajectories constrain only a low-dimensional projection of the parameterization. The framework therefore delivers a faithful behavioral emulator rather than a unique microscopic Hamiltonian, and the time-resolved validation quantifies where that emulator transfers: re-propagating the learned Hamiltonian with the full Lindblad equation reproduces the single-qubit dynamics with a trajectory-averaged fidelity of 0.94, while the multi-mode cases fall to about 0.58. A further contribution is the conversion of the learned open-system Hamiltonian into a physically interpretable two-qubit–bus-resonator circuit. The extracted parameters place qubit 1 firmly and qubit 2 marginally in the dispersive regime, and yield transmon capacitances of 77.3 fF, Josephson inductances of 9.2 and 11.2 nH, a 15.18 mm $\lambda/2$ bus resonator and a 7.59 mm qubit separation, all of which are compatible with planar superconducting fabrication; the reconstructed circuit reproduces the principal dynamical behavior of the reference systems. So, the results establish a practical pathway from black-box quantum-system identification to physical quantum-emulator and quantum-digital-twin construction, and they identify the three developments that would most improve it: a differentiable mixed-state propagator to lift the purity bound, a reduced or regularized operator basis to remove the structural degeneracies, and trajectory-conditioned network inputs together with experimental validation to establish genuine generalization across devices.

**Funding Declaration**: There is no funding support for this work.

**References**

[1] J. Preskill,"Quantum Computing in the NISQ era and beyond,"Quantum, vol. 2, p. 79, 2018.https://doi.org/10.22331/q-2018-08-06-79

[2] F. Arute et al.,"Quantum supremacy using a programmable superconducting processor,"Nature, vol. 574, pp. 505–510, 2019.https://doi.org/10.1038/s41586-019-1666-5

[3] C. L. Degen, F. Reinhard, and P. Cappellaro,"Quantum sensing,"Reviews of Modern Physics, vol. 89, 035002, 2017.https://doi.org/10.1103/RevModPhys.89.035002

[4] I. M. Georgescu, S. Ashhab, and F. Nori,"Quantum simulation,"Reviews of Modern Physics, vol. 86, pp. 153–185, 2014.https://doi.org/10.1103/RevModPhys.86.153
[5] H.-Y. Huang, Y. Liu, M. Broughton, I. Kim, A. Anshu, Z. Landau, and J. R. McClean, "Learning shallow quantum circuits," in Proc. 56th Annual ACM Symposium on Theory of Computing (STOC), ACM, 2024. https://doi.org/10.1145/3618260.3649722
[6] H.-Y. Huang, R. Kueng, and J. Preskill,"Predicting many properties of a quantum system from very few measurements,"Nature Physics, vol. 16, pp. 1050–1057, 2020.
[7] M. Paris and J. Řeháček (Editors),"Quantum State Estimation,"Springer, Lecture Notes in Physics, 2004.
[8] M. G. A. Paris,"Quantum estimation for quantum technology,"International Journal of Quantum Information, vol. 7, pp. 125–137, 2009.
[9] D. Hangleiter, I. Roth, J. Fuksa, J. Eisert, and P. Roushan, "Robustly learning the Hamiltonian dynamics of a superconducting quantum processor," Nature Communications, vol. 15, 9595, 2024.
[10] V. Gebhart, R. Santagati, A. A. Gentile, E. M. Gauger, D. Craig, N. Ares, L. Banchi, F. Marquardt, L. Pezzè, and C. Bonato, "Learning quantum systems," Nature Reviews Physics, vol. 5, pp. 141–156, 2023.
[11] M. Cerezo et al.,"Challenges and opportunities in quantum machine learning,"Nature Computational Science, vol. 2, pp. 567–576, 2022.
[12] V. Dunjko and H. J. Briegel,"Machine learning and artificial intelligence in the quantum domain," Reports on Progress in Physics, vol. 81, 074001, 2018.
[13] D. Burgarth and K. Yuasa,"Quantum system identification,"Physical Review Letters, vol. 108, 080502, 2012.
[14] C. Granade, C. Ferrie, and D. G. Cory,"Accelerated randomized benchmarking and Hamiltonian learning," New Journal of Physics, vol. 17, 013042, 2015.
[15] D. Dong and I. R. Petersen,"Quantum control theory and applications: A survey,"IET Control Theory & Applications, vol. 4, pp. 2651–2671, 2010.
[16] D'Ariano, G. M., Paris, M. G. A., & Sacchi, M. F., "Quantum Tomography," *Advances in Imaging and Electron Physics*, vol. 128, pp. 205–308, 2003.
[17] Nielsen, M. A., & Chuang, I. L., *Quantum Computation and Quantum Information*, Cambridge University Press, 2010.
[18] F. Huszár and N. M. T. Houlsby, "Adaptive Bayesian quantum tomography," Physical Review A, vol. 85, 052120, 2012. https://doi.org/10.1103/PhysRevA.85.052120
[19] A. Salmanogli and H. Zandi, "Implementing Grover algorithm on quantum chip architecture optimized with QGHNN for fidelity and entanglement preservation," arXiv:2511.04194 [quant-ph], 2026. https://doi.org/10.48550/arXiv.2511.04194
[20] Cerezo, M., Arrasmith, A., Babbush, R., *et al.*, "Variational Quantum Algorithms," *Nature Reviews Physics*, vol. 3, pp. 625–644, 2021.
[21] Koch, J., Yu, T. M., Gambetta, J., *et al.*, "Charge-Insensitive Qubit Design Derived from the Cooper Pair Box," *Physical Review A*, vol. 76, 042319, 2007.
[22] Glaser, S. J., Boscain, U., Calarco, T., *et al.*, "Training Schrödinger's Cat: Quantum Optimal Control," *European Physical Journal D*, vol. 69, 279, 2015.
[23] D. F. V. James, P. G. Kwiat, W. J. Munro, and A. G. White, Measurement of qubits, Phys. Rev. A 64, 052312 (2001).
[24] I. L. Chuang and M. A. Nielsen, Prescription for experimental determination of the dynamics of a quantum black box, J. Mod. Opt. 44, 2455–2467 (1997).

[25] S. T. Flammia and Y.-K. Liu, Direct Fidelity Estimation from Few Pauli Measurements, Phys. Rev. Lett. 106, 230501 (2011).
[26] R. Blume-Kohout, Robust Error Bars for Quantum Tomography, arXiv:1202.5270 (2012).
[27] D. Gross, Y.-K. Liu, S. T. Flammia, S. Becker, and J. Eisert, Quantum State Tomography via Compressed Sensing, Phys. Rev. Lett. 105, 150401 (2010).
[28] C. Ferrie, Self-Guided Quantum Tomography, Phys. Rev. Lett. 113, 190404 (2014).
[29] J. Carrasquilla and R. G. Melko, "Machine learning phases of matter," Nature Physics, vol. 13, pp. 431–434, 2017.
[30] A. M. Palmieri, E. Kovlakov, F. Bianchi, D. Yudin, S. Straupe, J. D. Biamonte, and S. Kulik, "Experimental neural network enhanced quantum tomography," npj Quantum Information, vol. 6, 20, 2020.
[31] W. Yu, J. Sun, Z. Han, and X. Yuan, "Robust and efficient Hamiltonian learning," Quantum, vol. 7, p. 1045, 2023. https://doi.org/10.22331/q-2023-06-29-1045
[32] A. Salmanogli, H. Zandi, Implementing Grover Algorithm on Quantum Chip Architecture Optimized with QGHNN for Fidelity and Entanglement Preservation, arXiv:2511.04194 [quant-ph], 2026.
[33] G. Torlai and R. G. Melko, "Machine-learning quantum states in the NISQ era," Annual Review of Condensed Matter Physics, vol. 11, pp. 325–344, 2020.
[34] Leng, J., Lin, S., Zhu, K., et al., *Differentiable Quantum Process Learning for Open Quantum Systems*, npj Quantum Information (2023).
[35] V. Bergholm, J. Izaac, M. Schuld, et al., "PennyLane: Automatic differentiation of hybrid quantum-classical computations," arXiv:1811.04968 [quant-ph], 2018. https://doi.org/10.48550/arXiv.1811.04968
[36] M. Schuld and F. Petruccione, *Machine Learning with Quantum Computers*, 2nd ed. (Springer, 2021).
[37] P. Rebentrost, M. Mohseni, and S. Lloyd, *Quantum Support Vector Machine for Big Data Classification*, Phys. Rev. Lett. 113, 130503 (2014).
[38] N. Wiebe, A. Kapoor, and K. M. Svore, "Quantum deep learning," Quantum Information & Computation, vol. 16, no. 7–8, pp. 541–587, 2016.
[39] S. Lloyd, *Universal Quantum Simulators*, Science 273, 1073–1078 (1996).
[40] M. Bukov, A. G. R. Day, D. Sels, P. Weinberg, A. Polkovnikov, and P. Mehta, "Reinforcement learning in different phases of quantum control," Physical Review X, vol. 8, 031086, 2018. https://doi.org/10.1103/PhysRevX.8.031086
[41] H.-Y. Huang, Y. Tong, D. Fang, and Y. Su, "Learning many-body Hamiltonians with Heisenberg-limited scaling," Physical Review Letters, vol. 130, 200403, 2023. https://doi.org/10.1103/PhysRevLett.130.200403
[42] Youssry, A., Yang, Y., Chapman, R.J. *et al.* Experimental graybox quantum system identification and control. *npj Quantum Inf* 10, 9 (2024). https://doi.org/10.1038/s41534-023-00795-5
[43] J. R. Johansson, P. D. Nation, and F. Nori, QuTiP 2: A Python framework for the dynamics of open quantum systems, Comput. Phys. Commun. 184, 1234 (2013). https://doi.org/10.1016/j.cpc.2012.11.019
[44] G. Torlai, G. Mazzola, J. Carrasquilla, M. Troyer, R. Melko, and G. Carleo, "Neural-network quantum state tomography," Nature Physics, vol. 14, pp. 447–450, 2018. https://doi.org/10.1038/s41567-018-0048-5.
[45] A. Paszke et al., *PyTorch: An Imperative Style, High-Performance Deep Learning Library*, NeurIPS, 2019.
[46] F. Schäfer, M. Kloc, C. Bruder, and N. Lörch, "A differentiable programming method for quantum control," Machine Learning: Science and Technology, vol. 1, 035009, 2020. https://doi.org/10.1088/2632-2153/ab9802.

[47] Wang, J., Paesani, S., Santagati, R. *et al.* Experimental quantum Hamiltonian learning. *Nature Phys* 13, 551–555 (2017). https://doi.org/10.1038/nphys4074.
[48] J. Watrous, *The Theory of Quantum Information*, Cambridge University Press, 2018.
[49] A. Uhlmann, "The transition probability in the state space of a *-algebra," *Reports on Mathematical Physics*, vol. 9, no. 2, pp. 273–279, 1976.
[50] R. Jozsa, "Fidelity for mixed quantum states," *Journal of Modern Optics*, vol. 41, no. 12, pp. 2315–2323, 1994.
[51] J. von Neumann, *Mathematical Foundations of Quantum Mechanics*, Princeton University Press, 1955.
[52] C. A. Fuchs and J. van de Graaf, "Cryptographic distinguishability measures for quantum-mechanical states," *IEEE Transactions on Information Theory*, vol. 45, no. 4, pp. 1216–1227, 1999.
[53] I. Bengtsson and K. Życzkowski, *Geometry of Quantum States*, 2nd Edition, Cambridge University Press, 2017.
[54] M. O. Scully and M. S. Zubairy, Quantum Optics. Cambridge, U.K.: Cambridge Univ. Press, 1997.
[55] A. Salmanogli, Quantum analysis of plasmonic coupling between quantum dots and nanoparticles, Phys. Rev. A 94, 043819, 2016
[56] A. Salmanogli, D. Gokcen, and H. S. Gecim, Entanglement of Optical and Microcavity Modes by Means of an Optoelectronic System, Phys. Rev. Applied 11, 024075, 2019.
[57] A. Salmanogli, Entangled microwave photons generation using cryogenic low noise amplifier (transistor nonlinearity effects), 2022 Quantum Sci. Technol. 7 045026, DOI 10.1088/2058-9565/ac8bf0.
[58] A. Mandviwalla, K. Ohshiro, and B. Ji, "Implementing Grover's algorithm on the IBM quantum computers," in Proc. IEEE Int. Conf. Big Data (Big Data), 2018, pp. 2531–2537, doi: 10.1109/BigData.2018.8622457.

[59] M. B. Plenio and P. L. Knight, "The quantum-jump approach to dissipative dynamics in quantum optics," Reviews of Modern Physics, vol. 70, no. 1, pp. 101–144, 1998. https://doi.org/10.1103/RevModPhys.70.101

[60] R. N. Gutenkunst, J. J. Waterfall, F. P. Casey, K. S. Brown, C. R. Myers, and J. P. Sethna, "Universally sloppy parameter sensitivities in systems biology models," PLoS Computational Biology, vol. 3, no. 10, e189, 2007. https://doi.org/10.1371/journal.pcbi.0030189
[61] M. A. Nielsen and I. L. Chuang, Quantum Computation and Quantum Information, 10th Anniversary Edition. Cambridge University Press, 2010.
[62] Y.-C. Liang, Y.-H. Yeh, P. E. M. F. Mendonça, R. Y. Teh, M. D. Reid, and P. D. Drummond, "Quantum fidelity measures for mixed states," Reports on Progress in Physics, vol. 82, 076001, 2019. https://doi.org/10.1088/1361-6633/ab1ca4
[63] C. W. Helstrom, Quantum Detection and Estimation Theory. New York: Academic Press, 1976.
[64] R. A. Horn and C. R. Johnson, Matrix Analysis, 2nd ed. Cambridge University Press, 2013.
[65] G. Lindblad, "On the generators of quantum dynamical semigroups," Communications in Mathematical Physics, vol. 48, no. 2, pp. 119–130, 1976.
[66] V. Gorini, A. Kossakowski, and E. C. G. Sudarshan, "Completely positive dynamical semigroups of N-level systems," Journal of Mathematical Physics, vol. 17, no. 5, pp. 821–825, 1976.
[67] H.-P. Breuer and F. Petruccione, The Theory of Open Quantum Systems. Oxford University Press, 2002.

[68] J. Dalibard, Y. Castin, and K. Mølmer, “Wave-function approach to dissipative processes in quantum optics,” Physical Review Letters, vol. 68, no. 5, pp. 580–583, 1992.
[69] K. Mølmer, Y. Castin, and J. Dalibard, “Monte Carlo wave-function method in quantum optics,” Journal of the Optical Society of America B, vol. 10, no. 3, pp. 524–538, 1993.
[70] A. J. Daley, “Quantum trajectories and open many-body quantum systems,” Advances in Physics, vol. 63, no. 2, pp. 77–149, 2014. https://doi.org/10.1080/00018732.2014.933502
[71] H. M. Wiseman and G. J. Milburn, Quantum Measurement and Control. Cambridge University Press, 2010.
[72] S. Blanes, F. Casas, J. A. Oteo, and J. Ros, “The Magnus expansion and some of its applications,” Physics Reports, vol. 470, no. 5–6, pp. 151–238, 2009. https://doi.org/10.1016/j.physrep.2008.11.001
[73] S. Amari, “Natural gradient works efficiently in learning,” Neural Computation, vol. 10, no. 2, pp. 251–276, 1998.
[74] J. Martens, “New insights and perspectives on the natural gradient method,” Journal of Machine Learning Research, vol. 21, no. 146, pp. 1–76, 2020.
[75] J. Nocedal and S. J. Wright, Numerical Optimization, 2nd ed. New York: Springer, 2006.
[76] S. L. Braunstein and C. M. Caves, “Statistical distance and the geometry of quantum states,” Physical Review Letters, vol. 72, no. 22, pp. 3439–3443, 1994.
[77] J. Liu, H. Yuan, X.-M. Lu, and X. Wang, “Quantum Fisher information matrix and multiparameter estimation,” Journal of Physics A: Mathematical and Theoretical, vol. 53, 023001, 2020. https://doi.org/10.1088/1751-8121/ab5d4d
[78] M. K. Transtrum, B. B. Machta, and J. P. Sethna, “Geometry of nonlinear least squares with applications to sloppy models and optimization,” Physical Review E, vol. 83, 036701, 2011. https://doi.org/10.1103/PhysRevE.83.036701
[79] M. K. Transtrum, B. B. Machta, K. S. Brown, B. C. Daniels, C. R. Myers, and J. P. Sethna, “Perspective: Sloppiness and emergent theories in physics, biology, and beyond,” The Journal of Chemical Physics, vol. 143, 010901, 2015. https://doi.org/10.1063/1.4923066
[80] A. Raue, C. Kreutz, T. Maiwald, J. Bachmann, M. Schilling, U. Klingmüller, and J. Timmer, “Structural and practical identifiability analysis of partially observed dynamical models by exploiting the profile likelihood,” Bioinformatics, vol. 25, no. 15, pp. 1923–1929, 2009. https://doi.org/10.1093/bioinformatics/btp358
[81] E. Bairey, I. Arad, and N. H. Lindner, “Learning a local Hamiltonian from local measurements,” Physical Review Letters, vol. 122, 020504, 2019. https://doi.org/10.1103/PhysRevLett.122.020504
[82] A. Gu, L. Cincio, and P. J. Coles, “Practical Hamiltonian learning with unitary dynamics and Gibbs states,” Nature Communications, vol. 15, 312, 2024. https://doi.org/10.1038/s41467-023-44008-1
[83] C. Spearman, “The proof and measurement of association between two things,” The American Journal of Psychology, vol. 15, no. 1, pp. 72–101, 1904.
[84] A. Blais, A. L. Grimsmo, S. M. Girvin, and A. Wallraff, “Circuit quantum electrodynamics,” Reviews of Modern Physics, vol. 93, 025005, 2021. https://doi.org/10.1103/RevModPhys.93.025005
[85] J. Koch, T. M. Yu, J. Gambetta, A. A. Houck, D. I. Schuster, J. Majer, A. Blais, M. H. Devoret, S. M. Girvin, and R. J. Schoelkopf, “Charge-insensitive qubit design derived from the Cooper pair box,” Physical Review A, vol. 76, 042319, 2007.

## Appendices

To evaluate the performance of the proposed QNN for Hamiltonian learning rigorously, complementary quantum-state validation metrics are employed. Since no single metric can completely characterize the agreement between two quantum states, two well-established quantities from quantum information theory are considered alongside the trajectory-density loss: the quantum-state fidelity and the trace distance. Each captures a different physical aspect of the reconstructed density matrix, namely the state overlap and the operational distinguishability, and together they provide a compact assessment of whether the reconstructed Hamiltonian reproduces the dynamics of the unknown open quantum system. Similar validation methodologies are widely adopted in quantum tomography, quantum computing, quantum control and Hamiltonian-learning studies [23, 48–53]. Appendices A and B define the two metrics and record the elementary properties of them that are used elsewhere in this work. Appendices C–F then derive the structural results that Secs. II and III invoke but do not establish: the bounds that the purity of the target state imposes on both metrics whenever the predicted state is pure (Appendix C); the exact relation between the differentiable propagator of Eq. 5 and the Lindblad generator of Eq. 1, together with the approximation that separates them (Appendix D); the exact effect of the chirped modulation $\alpha(t)$ on the learned dynamics (Appendix E); and the exact rank deficiency of the thirty-two-operator basis (Appendix F). Together these four results account for the ceiling on the state-level metrics reported in Table III, for the role of the chirp identified in Fig. 5f, and for the operator-level degeneracy that underlies the central negative result of Fig. 9.

***Appendix A. Quantum State Fidelity***

Among all similarity measures, **quantum state fidelity** is arguably the most widely adopted metric for comparing quantum states. Fidelity measures the overlap between two density matrices and quantifies the probability that one state can successfully reproduce the other. For two arbitrary mixed quantum states, $\rho_1$ and $\rho_2$, the Uhlmann fidelity is defined as:

$$F(\rho_1, \rho_2) = \left[\mathrm{Tr}\left(\sqrt{\sqrt{\rho_1}\rho_2\sqrt{\rho_1}}\right)\right]^2. \tag{A1}$$

where the matrix square-root operation is used to compute. The fidelity satisfies as:

$$0 \le F \le 1, \tag{A2}$$

where $F = 1$ indicates identical quantum states and $F = 0$ corresponds to orthogonal states. Owing to its invariance under unitary transformations and clear operational interpretation, fidelity has become one of the standard validation metrics in quantum state tomography, quantum control, quantum communication, quantum computing, and Hamiltonian identification [47–50]. In the present work, fidelity serves as the primary indicator of how accurately the reconstructed Hamiltonian reproduces the dynamics of the unknown quantum system. One property of Eq. A1 is used repeatedly in this work and is worth stating separately, namely its form when one of the two states is pure. If $\rho_1 = |\psi\rangle\langle\psi|$, then $\rho_1$ is its own square root, and the operator appearing inside the trace of Eq. A1 becomes $\sqrt{\rho_1}\, \rho_2\, \sqrt{\rho_1} = \langle\psi|\rho_2|\psi\rangle\, |\psi\rangle\langle\psi|$, which is a non-negative multiple of a one-dimensional projector. Its square root is $\langle\psi|\rho_2|\psi\rangle$^(1/2) $|\psi\rangle\langle\psi|$, so the trace equals $\langle\psi|\rho_2|\psi\rangle$^(1/2) and Eq. A1 reduces to

$$F(|\psi\rangle\langle\psi|, \rho_2) = \langle\psi|\rho_2|\psi\rangle \tag{A3}$$

The Uhlmann fidelity then coincides with the expectation value of the mixed state in the pure state, which is the elementary form used in Appendix C. This is the situation encountered throughout the present work, because the differentiable propagator of Eq. 5 returns a pure state at every time step, for the reason derived in Appendix D.

### *Appendix B. Trace Distance*

Another fundamental metric employed in this work is the **trace distance**, which measures how distinguishable two quantum states are.

For two density matrices, $\rho_1, \rho_2$, the trace distance is:

$$D(\rho_1, \rho_2) = \frac{1}{2}\mathrm{Tr} \mid \rho_1 - \rho_2 \mid. \tag{B1}$$

By considering $\Delta = \rho_1 - \rho_2$. The **trace distance is calculated as:**

$$D = \frac{1}{2}\sum_i \mid \lambda_i \mid, \tag{B2}$$

where $\lambda_i$are the eigenvalues of the Hermitian matrix $\Delta$. The trace distance satisfies by:

$$0 \leq D \leq 1, \tag{B3}$$

where $D = 0$ indicates identical quantum states and $D = 1$ denotes perfectly distinguishable states. Unlike fidelity, trace distance directly quantifies state distinguishability and possesses an operational interpretation as the maximum probability of distinguishing two quantum states through optimal measurements. Therefore, fidelity and trace distance provide complementary measures for evaluating Hamiltonian reconstruction performance [47,48,52]. For the derivations of Appendix C it is convenient to record also the variational form of Eq. B1. For any Hermitian operator $\Delta$ the trace norm satisfies $\frac{1}{2}\mathrm{Tr}|\Delta| = \max \mathrm{Tr}[P\Delta]$, the maximum being taken over all operators obeying $0 \leq P \leq I$ and being attained by the projector onto the positive part of $\Delta$. Hence

$$D(\rho_1, \rho_2) = \max_{0 \leq P \leq I} \mathrm{Tr}[P(\rho_1 - \rho_2)] \tag{B4}$$

which is the precise sense in which the trace distance is the largest bias achievable with a single binary measurement, and which is the origin of the operational interpretation quoted above [48, 52]. Any particular choice of P therefore yields a valid lower bound on D, a fact used in Eq. C2.

### *Appendix C. Attainability Bounds for a Pure Predicted State*

The differentiable propagator used for training returns a pure state at every time step, $\rho_{\mathrm{pred}}(t) = |\psi(t)\rangle\langle\psi(t)|$, for the reason derived in Appendix D, whereas the target trajectory produced by the full Lindblad solver becomes progressively mixed as dissipation proceeds. This asymmetry imposes a hard ceiling on both validation metrics, which has to be taken into account before the values reported in Table III are interpreted. Writing $\lambda_{\mathrm{max}}(\rho_{\mathrm{unk}})$ for the largest eigenvalue of the target density matrix and using Eq. A3, the fidelity obeys

$$F(\rho_{\mathrm{pred}}, \rho_{\mathrm{unk}}) = \langle\psi|\rho_{\mathrm{unk}}|\psi\rangle \leq \lambda_{\mathrm{max}}(\rho_{\mathrm{unk}}) \tag{C1}$$

because $\langle\psi|\rho_{\mathrm{unk}}|\psi\rangle$ is a normalized expectation value of a positive operator and is therefore bounded by its largest eigenvalue, with equality if and only if $|\psi\rangle$ is a principal eigenvector of $\rho_{\mathrm{unk}}$. The trace distance is bounded from below by the same quantity: the choice $P = |\psi\rangle\langle\psi|$ in Eq. B4 gives

$$D(\rho_{\mathrm{pred}}, \rho_{\mathrm{unk}}) \geq \langle\psi|(\rho_{\mathrm{pred}} - \rho_{\mathrm{unk}})|\psi\rangle = 1 - F \geq 1 - \lambda_{\mathrm{max}}(\rho_{\mathrm{unk}}) \tag{C2}$$

Both bounds depend on the target state alone and are entirely independent of the quality of the learned Hamiltonian. They are also never vacuous, since the purity of the target provides a lower bound on $\lambda_{\mathrm{max}}$,

$$\mathrm{Tr}\rho_{\mathrm{unk}}^2 = \sum_i {\lambda_i}^2 \leq \lambda_{\mathrm{max}} \sum_i \lambda_i = \lambda_{\mathrm{max}}(\rho_{\mathrm{unk}}) \tag{C3}$$

so that the attainable fidelity is at least as large as the purity of the state being reconstructed. Eqs C1 and C2 are the bounds evaluated for every run in Sec. III and drawn as the limiting lines in Fig. 8(a) and Fig. 8(b). A complementary question is whether the bound of Eq. C1 is approachable in practice, or whether the no-jump construction additionally misaligns the predicted state with the principal eigenvector of the target.

This can be answered directly by propagating the exact unknown Hamiltonian through Eq. 5 and comparing the resulting state with the principal eigenvector of the corresponding Lindblad solution. The squared overlap $|\langle\psi|v_{max}\rangle|^2$ obtained in this way has a trajectory average of 0.990 for $H_1$, 0.997 for $H_2$ and 0.991 for the genuinely unknown system, rising to between 0.99 and 1.00 at the end of the evolution window. The direction of the predicted state is therefore essentially correct even though its rank is not, and an exact Hamiltonian estimate would recover approximately 0.99 $\lambda_{max}$. The fidelity deficit associated with the propagator is consequently structural rather than a failure of estimation, and the normalized quantity $F/\lambda_{max}$ reported in Fig. 8c and Table IV is the appropriate figure of merit: twelve of the eighteen runs exceed 0.90 and the best attains 1.00, whereas the raw fidelities of Table III range from 0.35 to 0.97.

***Appendix D. The Differentiable Propagator and the Lindblad Generator***

Equation 5 is not a discretization of Eq. 1, and the difference is material to every result reported in Sec. III. Collecting the anticommutator part of Eq. 2 into a non-Hermitian generator, the Lindblad equation can be written exactly as

$$\frac{d\rho}{dt} = -i\left(H_{\mathrm{eff}}\rho - \rho H_{\mathrm{eff}}^{\dagger}\right) + \sum_{k} L_k \rho L_k^{\dagger} \quad \text{(D1)}$$

with

$$H_{\mathrm{eff}} = H - \frac{i}{2}\sum_{k} L_k^{\dagger} L_k \quad \text{(D2)}$$

The first term of Eq. D1 describes the deterministic, norm-non-preserving damping of the state, while the second, commonly called the recycling or jump term, restores the population that the first term removes. Discarding the recycling term leaves a linear equation whose solution is available in closed form,

$$\tilde{\rho}(t) = e^{-iH_{\mathrm{eff}}t}\rho(0)e^{+iH_{\mathrm{eff}}^{\dagger}t} \quad \text{(D3)}$$

and the normalized state that Eq. 5 propagates is

$$\rho_{\mathrm{pred}}(t) = \frac{\tilde{\rho}(t)}{\mathrm{Tr}\tilde{\rho}(t)} \quad \text{(D4)}$$

Three consequences follow. First, Tr $\tilde{\rho}(t)$ is the probability that no quantum jump has occurred up to time t, so $\rho_{pred}(t)$ is not an approximation to the Lindblad state at all, but the exact state of the system conditioned on no jump having been detected; Eq. 5 is the deterministic part of a quantum-trajectory unravelling of Eq. 1 [59]. Second, Eq. D3 is a congruence transformation and therefore preserves rank, so an initially pure state remains rank one at all times and the renormalization of Eq. D4 keeps it normalized. This is the origin of the purity constraint used in Appendix C and displayed directly in Fig. 10c. Third, the accuracy of the construction is governed by the magnitude of the discarded term relative to the retained one, that is by the jump probability $1 - \mathrm{Tr}\,\tilde{\rho}(t)$. For the three systems examined in Fig. 10 the survival probability Tr $\tilde{\rho}(T)$ at the end of the 50 ns window is 0.57, 0.55 and 0.50 respectively, so that approximately half of the ensemble has undergone a jump by the end of the trajectory. This is precisely why the target states are strongly mixed, with $\lambda_{max}$ falling to between 0.79 and 0.94, and it quantifies how much of the residual reported in Table III is attributable to the propagator rather than to the learned Hamiltonian.

A separate point concerns the time discretization. Because Heff is time-independent and α(t) enters only as a scalar perfector, the matrix exponential in Eq. 5 is exact within each step and no Trotter or commutator error is incurred; for constant α the propagator is exact for any step size. The only discretization error is the piecewise-constant sampling of α(t) itself, which displaces the accumulated effective time defined in Eq. E2. For the representative sinusoidal chirp on the 200-point grid used throughout this work that displacement amounts to 2.8% of the total accumulated effective time, and for constant propagation it vanishes identically.

***Appendix E. Exact Effect of the Chirped Modulation***

The role of the modulation α(t) introduced in Eq. 5 can be established exactly, without approximation. Since Heff does not depend on time and α(t) is a scalar, the generators $\alpha(t_1)H_{\mathrm{eff}}$ and $\alpha(t_2)H_{\mathrm{eff}}$ commute for every pair of times, and the ordered product of the step propagators collapses to a single exponential,

$$|\psi(t)\rangle \propto \exp[-iH_{\mathrm{eff}}s(t)]|\psi(0)\rangle \tag{E1}$$

where

$$s(t) = \int_0^t \alpha(t')dt' \tag{E2}$$

up to the normalization of Eq. D4. Evaluating both sides at the converged coefficients of the unknown system reproduces the stepwise propagator of Eq. 5 to $4\times10^{-16}$ in state overlap, that is to machine precision. The chirped trajectory is therefore not a new trajectory at all: it is the unmodulated trajectory of the same Hamiltonian, read on a deformed time axis s(t).

Three consequences follow, and together they account for the behavior reported in Fig. 5d–f. First, the modulated and unmodulated predictions are related by a reparameterization of time that does not involve θ, and α(t) never enters the generation of the target trajectories; the modulation therefore cannot increase the information that the data carry about the unknown Hamiltonian. The statement made in Sec. II on physical grounds is thus exact rather than heuristic. Second, the null space of the Gauss–Newton metric is independent of α, because a coefficient combination that leaves Heff unchanged leaves the right-hand side of Eq. E1 unchanged for every value of s; the three exactly null directions established in Appendix F consequently persist under all three propagation modes, as observed in Fig. 5d. Third, although the null space is invariant, the curvature is not. Expanding Eq. E1 to first order in s gives $\partial\rho(t)/\partial\theta_k \approx -i\, s(t)[B_k, \rho(0)]$, so that the dominant eigenvalue of the Gauss–Newton metric scales as the weighted mean square of the accumulated effective time,

$$\frac{\lambda_1(\alpha)}{\lambda_1(1)} \approx \frac{\langle s^2\rangle_\alpha}{\langle s^2\rangle_1} \tag{E3}$$

with

$$\langle s^2\rangle = \frac{1}{N}\sum_{j=1}^{N} w_j s^2(t_j) \tag{E4}$$

For the Gaussian profile this ratio evaluates to 0.18 against a measured 0.23, and for the sinusoidal chirp to 0.020 against a measured 0.070; the ordering and the order of magnitude are reproduced, and the residual discrepancy is the expected breakdown of the first-order expansion over a 50 ns window. The mechanism is made concrete by the accumulated effective time itself. For the representative sinusoidal chirp s(t) is non-monotonic and spans only 7.5 ns, and for the Gaussian profile 18.8 ns, against the 50 ns of constant propagation. The modulation compresses the effective propagation interval and flattens the loss surface, thereby accelerating convergence, while leaving the information content of the data untouched. This is the precise sense in which the chirp acts as an implicit preconditioner of the optimization rather than as an excitation applied to the unknown device, and it explains why chirped modulation improves the trajectory loss of Figs. 2–4 without improving the state-level fidelity of Table III.

***Appendix F. Exact Rank Deficiency of the Operator Basis***

The parameterization of Eq. 3 assigns one coefficient to each of thirty-two basis operators, but the map $\theta \mapsto H_{unknown}$ is not injective. Two exact operator identities are responsible. The first is the standard relation between the exchange interaction and the transverse two-qubit couplings,

$$\sigma_+^{(1)}\sigma_-^{(2)} + \sigma_-^{(1)}\sigma_+^{(2)} = \frac{1}{2}\left(\sigma_x^{(1)}\sigma_x^{(2)} + \sigma_y^{(1)}\sigma_y^{(2)}\right) \quad \text{(F1)}$$

which makes the exchange operator of the basis an exact linear combination of the XX and YY operators. The second follows from the truncation of both bosonic modes to their two lowest levels. On a two-dimensional subspace the number operator is the projector onto the excited level, so that

$$n = |1\rangle\langle 1|, \quad n^2 = n \quad \text{(F2)}$$

and the Kerr-type operators $na^2$ and $nb^2$ coincide identically with na and nb. The basis therefore contains three exact linear dependencies. This is confirmed by the Hilbert–Schmidt Gram matrix

$$G_{kl} = \mathrm{Tr}\left[B_k^\dagger B_l\right] \quad \text{(F3)}$$

whose rank is twenty-nine rather than thirty-two; the three vanishing eigenvectors ck satisfy $||\Sigma_k c_k B_k||_F \approx 3\times10^{-15}$, that is, they vanish to numerical precision. The consequence for the identification problem is structural rather than numerical. Writing K for the three-dimensional kernel of the map $\theta \mapsto H_{unknown}$, every coefficient vector in the coset $\theta + K$ produces exactly the same Hamiltonian,

$$H(\theta + c) = H(\theta) \quad for\ all\ c \in K \quad \text{(F4)}$$

and hence exactly the same trajectory, the same measurement statistics and the same value of the loss of Eq. 7. The maximum-likelihood problem therefore possesses an affine solution set of dimensions at least three, and no increase in the quantity, the duration or the precision of the data can reduce it. Only a change of parameterization can do so: removing the redundant operators, or equivalently reporting the estimate in the twenty-nine-dimensional quotient space, restores a well-posed problem at the level of principle.

It should be stated, however, that this exact degeneracy is not the dominant obstacle in practice. Projecting both the true and the learned coefficient vectors onto the identifiable twenty-nine-dimensional subspace changes the correlation between them only marginally, from −0.25 to −0.28 for the sinusoidally chirped run on the unknown system and from −0.14 to −0.22 for the Gaussian one, and leaves the relative coefficient error of order unity in every case. The quantitative failure of parameter recovery documented in Fig. 9 is therefore governed by the approximately ten numerically flat but non-null directions visible in Fig. 5d rather than by the three exactly null ones. In this respect the present inverse problem displays the behavior characteristic of sloppy models, in which the eigenvalues of the parameter-sensitivity metric are spread roughly uniformly over many decades and only a few stiff collective combinations are constrained by the data [60].

### *Appendix G: Gauss–Newton (Fisher) metric and technical information about stiffest, second-stiffest, sloppiest identifiable and an exactly null eigendirection*

The identifiability statements made in connection with Fig. 5d–Fig. 5f rest on the local curvature of the trajectory loss at the converged coefficients. This appendix defines that object precisely, states how it is evaluated, and reports the composition of the four eigendirections that the main text refers to but does not describe.

#### *G.1 Definition and normalization*

For a nonlinear least-squares objective the Hessian separates into a term quadratic in the model Jacobian and a term proportional to the residual [75]. Discarding the second, which is the Gauss–Newton approximation and is accurate near a good fit, the curvature of the weighted trajectory loss is governed by

$$G_{kl} = \frac{1}{N}\sum_{j=1}^{N} w_j \,\mathrm{Re}\,\mathrm{Tr}\left[\left(\frac{\partial\rho(t_j)}{\partial\theta_k}\right)^{\dagger}\frac{\partial\rho(t_j)}{\partial\theta_l}\right],\quad w_j = j/N \tag{G1}$$

which is a Fisher metric on the thirty-two-dimensional coefficient manifold in the sense of Amari [73] and coincides with the generalized Gauss–Newton matrix for this class of objective [74]. It is a classical Fisher metric of the trajectory residual and should not be confused with the quantum Fisher information of the state, which bounds the precision attainable by an optimal measurement [76, 77]; the two answer different questions, the former about the conditioning of this estimator and the latter about the ultimate information content of the state.

Equation G1 uses the full Hilbert–Schmidt trace, whereas the loss of Eq. 7 is normalized by the $D^2 = 256$ matrix elements. The two are related by

$$\frac{\partial^2\mathcal{L}}{\partial\theta_k\partial\theta_l} \approx \frac{2G_{kl}}{D^2} \tag{G2}$$

so that a displacement $\delta\theta = \delta\, v_i$ along the $i^{th}$ eigendirection changes the loss by

$$\Delta\mathcal{L} \approx \frac{\lambda_i\delta^2}{D^2} \tag{G3}$$

All statements in the main text concern ratios of eigenvalues and are therefore independent of this convention; Eq. G3 is required only when the spectrum is compared with an absolute loss value, as in G.4 below.

### *G.2 Numerical evaluation*

The derivatives in Eq. G1 are evaluated by central finite differences with step $h = 10^{-4} \times 2\pi$ GHz, applied one coefficient at a time to the converged vector of the thirty-two-coefficient unknown system, and propagated with the same differentiable propagator of Eq. 5 that is used for training, so that the metric describes the landscape the optimizer actually sees. The step is four orders of magnitude below the output range of the network and four orders above the level at which round-off in the propagator becomes visible. For the chirped cases the metric is averaged over five independent draws of the chirp parameters from the ranges given in Sec. II, because a single draw fixes one particular accumulated effective time and is not representative, as shown in Appendix E. The resulting spectra are summarized in Table G1.

Table G1. Spectrum of the Gauss–Newton metric of Eq. G1 at the converged coefficients of the unknown system, for constant propagation and for the sinusoidal chirp averaged over five draws. Eigenvalues are given in the normalization of Eq. G1, with the coefficients measured in units of 2π GHz.

| **Quantity** | **Constant drive** | **Sinusoidal chirp** |
|---|---|---|
| $\lambda_1$ (dominant curvature) | $1.257\times10^{3}$ | $8.83\times10^{1}$ |
| $\lambda_2$ | $6.74\times10^{2}$ | $2.20\times10^{1}$ |
| $\lambda_{20}$ | $6.8\times10^{-3}$ | $3.9\times10^{-4}$ |
| $\lambda_{29}$ (sloppiest identifiable) | $1.30\times10^{-5}$ | $1.07\times10^{-5}$ |
| $\lambda_{30}$, $\lambda_{31}$, $\lambda_{32}$ (null space) | $1.2\times10^{-10}$, $1.2\times10^{-11}$, $1.0\times10^{-12}$ | $2.7\times10^{-13}$, $1.0\times10^{-14}$, $9.5\times10^{-16}$ |
| Tr G | $2.47\times10^{3}$ | $1.41\times10^{2}$ |
| $\log_{10}(\lambda_1/\lambda_{29})$ | 7.99 | 6.92 |
| Participation ratio | 2.83 | 2.31 |
| Numerical rank ($10^{-6}\,\lambda_1$) | 20 | 22 |

The last three eigenvalues are zero to machine precision in both cases; the non-zero values quoted are the accumulated round-off of the finite-difference evaluation, and they fall by three further orders of magnitude under the chirp only because the overall scale of the metric is smaller. They correspond to the exact null space established in Appendix F. The participation ratio, the square of Tr G divided by the sum of the squared eigenvalues, is close to two in both cases: essentially all of the curvature is carried by two collective directions out of thirty-two, which is the quantitative form of the sloppiness described in the main text and is the behavior catalogued for models of this kind in [78, 79].

### *G.3 Composition of the four eigendirections*

The directions referred to in Fig. 5e are not abstract: each is dominated by a small group of basis operators with a clear physical reading. Table G2 lists the leading components for the sinusoidal-chirp metric, which is the case plotted in Fig. 5e. In every case the five largest components carry at least 99% of the squared weight, so the directions are genuinely sparse in the operator basis rather than delocalized over it.

Table G2. Leading components of the eigendirections of the Gauss–Newton metric for the sinusoidal chirp. The coefficients are components of the unit eigenvector in the thirty-two-operator basis of Eq. 3.

| Direction | λ | Leading components | Physical content |
|---|---|---|---|
| v1 stiffest | $8.83\times10^{1}$ | na (+0.63), $\sigma z^1$na (−0.51), $\sigma z^2$na (+0.51), Pa (−0.23), Xa (+0.17) | Resonator population together with the two dispersive shifts – the readout channel |
| v2 second stiffest | $2.20\times10^{1}$ | nb (−0.66), $\sigma z^2$nb (−0.53), $\sigma z^1$nb (+0.53), na (+0.04) | The same combination for the bus mode |
| v29 sloppiest identifiable | $1.07\times10^{-5}$ | $\sigma x^1$na (−0.68), $\sigma y^1$Xa (+0.60), $\sigma x^1$Xa (+0.41), $\sigma x^2$na (−0.06) | Transverse qubit–resonator couplings, pre-weighted by 0.01–0.02 and rapidly rotating in the dispersive frame |
| v32 exact null | $9.5\times10^{-16}$ | $\sigma+^1\sigma-^2$ (+0.75), $\sigma y^1\sigma y^2$ (−0.45), $\sigma x^1\sigma x^2$ (−0.45), $nb^2$ (+0.21) | The exchange–XX–YY dependency of Eq. F1; unobservable by construction |

Three observations follow. First, the stiff directions are exactly the dispersive qubit–resonator sector, which is both the physically dominant interaction in the sampled Hamiltonians and the one most strongly imprinted on the resonator observables that dominate the trajectory. The optimization is therefore well conditioned precisely where the physics is strong, which is reassuring but also explains why little is learned about anything else. Second, the sloppiest identifiable direction consists of the transverse light–matter terms that the physics-informed weighting of Sec. II deliberately suppresses by factors of 0.01 to 0.02; their effect on the trajectory is smaller than that suppression alone would suggest, because in the dispersive regime they rotate rapidly and average out over the 50 ns window. The apparent sloppiness is thus a physical statement about what a dispersive measurement can resolve, not an artefact of the parameterization. Third, the exact null direction recovered numerically is (+0.75, −0.45, −0.45) on the exchange, YY and XX operators, which is the normalized form of the identity of Eq. F1 and confirms Appendix F independently.

### *G.4 Loss response and the limits of the quadratic description*

Fig. 5e reports the change in the trajectory loss when the converged coefficients are displaced along these directions. Table G3 compares the measured response with the second-order prediction of Eq. G3.

Table G3. Measured (synthetically) and predicted change in the trajectory loss for a displacement δ along each eigendirection, in units of 2π GHz. The base loss is $\mathscr{L}(\theta) = 7.87\times10^{-4}$.

| Direction | δ = 0.01 | δ = 0.05 | δ = 1.00 | $\Delta\mathscr{L}/\mathscr{L}$ at δ = 1 |
|---|---|---|---|---|
| v1 measured | $+2.7\times10^{-5}$ | $+4.8\times10^{-4}$ | $+8.7\times10^{-4}$ | +1.10 |
| v1 predicted, Eq. G3 | $+3.4\times10^{-5}$ | $+8.6\times10^{-4}$ | $+3.4\times10^{-1}$ | — |
| v2 measured | $-2.1\times10^{-6}$ | $+4.4\times10^{-5}$ | $+1.3\times10^{-3}$ | +1.67 |
| v29 measured | $-8.1\times10^{-10}$ | $-4.1\times10^{-9}$ | $-1.2\times10^{-7}$ | $-1.46\times10^{-4}$ |
| v32 measured | $-2.5\times10^{-14}$ | $-1.3\times10^{-13}$ | $-2.5\times10^{-12}$ | $-3.21\times10^{-9}$ |

The quadratic description is accurate only where it should be. Along the stiffest direction it reproduces the measured response to within 20% at $\delta = 0.01$ and then fails progressively, as expected once the displacement becomes comparable to the curvature radius of the model manifold [78]. Along $v_{29}$ and $v_{32}$ the measured change is negative and scales linearly rather than quadratically in $\delta$, which is the signature of a residual gradient rather than of curvature: the optimizer has not reached stationarity along these directions, because the curvature that would drive it there is smaller than the gradient noise of the stochastic updates. This is not a defect of the training but the expected outcome for a sloppy objective, and it is the practical reason why the coefficient estimate is non-unique even though the trajectory is reproduced well.

For the exact null direction the measured change, of order $10^{-12}$ in relative terms, is pure floating-point round-off: the displacement changes neither the Hamiltonian nor any observable, as Eq. F4 requires. A full displacement of $2\pi \times 1$ GHz, the entire output range of the network, along this direction leaves the loss unchanged to nine significant figures.

### *G.5 Reading the spectrum as an identifiability statement*

The spectrum of Eq. G1 supports the same conclusion as a profile-likelihood analysis would, and in the same language [80]. A direction is practically identifiable only if a displacement along it changes the loss by more than the level at which the loss itself is meaningful, which here is set by the injected measurement noise, $\sigma^2 \approx 5\times10^{-8}$. Applying this criterion to Table G3 places the boundary between the identifiable and the unidentifiable part of the coefficient space near $v_{20}$: the twenty directions above $10^{-6}$ of $\lambda_1$ produce loss changes above the noise floor over the accessible displacement range, whereas the remaining nine identifiable-but-flat directions and the three null ones do not. The thirty-two-coefficient parameterization therefore carries at most about twenty effective degrees of freedom under the present measurement protocol, which is the quantitative version of the statement that the framework delivers a behavioral emulator rather than a unique Hamiltonian.

It should be stressed that this boundary is a property of the measurement protocol and not of the network. Lengthening the trajectory, probing from initial states that populate more of the Hilbert space, or measuring observables that couple to the transverse sector would all move it. Estimating how far it can be moved, and at what experimental cost, is the natural continuation of this work.